\documentclass[a4paper,USenglish,cleveref]{lipics-v2021}
\usepackage[refchapterCite,hideChapterNumber,noCCBy]{craigstyle}

\usepackage{xspace}
\usepackage[usenames,dvipsnames]{xcolor} 
\usepackage{tikz}
\usepackage{stmaryrd} %
\usetikzlibrary{shapes.geometric}
\usetikzlibrary{shapes.symbols}
\usetikzlibrary{positioning}
\usetikzlibrary{arrows}
\usepackage{placeins}
\usepackage{csquotes}
\usepackage{examplesyms}

\newcommand{\txtcomment}[3]
{{\color{#2}{\textbf{#1 (before March 2025):} #3}}}

\newcommand{\txtcommentNew}[3]
{{\color{#2}{\textbf{#1 (after May 2025):} #3}}}

\newcommand{\red}[1]{{\color{red} #1}}

\renewcommand{\txtcomment}[3]{}
\renewcommand{\txtcommentNew}[3]{}
\renewcommand{\red}[1]{#1}

\newcommand{\sig}{\mathit{sig}}
\newcommand{\sub}{\mathit{sub}}
\newcommand{\sigposneg}{\mathit{sig}^\pm}
\newcommand{\complit}[1]{\overline{#1}}

\newcommand{\defname}[1]{\emph{#1}} %
\newcommand{\name}[1]{\emph{#1}} %
\newcommand{\entails}{\models}

\newcommand{\imp}{\rightarrow}
\newcommand{\equi}{\leftrightarrow}
\newcommand{\true}{\top}
\newcommand{\false}{\bot}
\newcommand{\tup}[1]{(#1)}

\DeclareMathOperator{\tabsep}{\;\raisebox{-1.5pt}{$\stackrel{\mathit{sep}}{\rightarrow}$}\;}

\title{\vspace{26pt}Interpolation in Classical Propositional Logic}

\author{Patrick Koopmann}
       {Vrije Universiteit Amsterdam, The Netherlands}
       {p.k.koopmann@vu.nl}
       {https://orcid.org/0000-0001-5999-2583}
       {}
\author{Christoph Wernhard}
       {University of Potsdam, Germany}
       {info@christophwernhard.com}
       {https://orcid.org/0000-0002-0438-8829}
       {}
\author{Frank Wolter}
       {University of Liverpool, UK}
       {wolter@liverpool.ac.uk}
       {https://orcid.org/0000-0002-4470-606X}
       {}

\authorrunning{P. Koopmann, C. Wernhard, F. Wolter}

\allowdisplaybreaks
\begin{document}

\maketitle

\begin{abstract}
 We introduce Craig interpolation and related notions such as uniform interpolation, Beth definability, and theory decomposition in classical propositional logic. We present four approaches to computing interpolants: via quantifier elimination, from formulas in disjunctive normal form, and by extraction from resolution or tableau refutations. We close with a discussion of the size of interpolants and links to circuit complexity.  
\end{abstract}

\tableofcontents

\section{Introduction}
In this chapter, we introduce Craig interpolation in the context of classical propositional logic. We aim to cover the fundamental results and techniques in the field, while still being accessible to readers who have only very basic logic background. We focus on explaining the underpinning logical notions and algorithms. For a discussion of applications of Craig interpolation in areas such as verification, databases, knowledge representation, and philosophy we refer the reader to the respective chapters of this volume.

In detail, we start by introducing the notion of a Craig interpolant between formulas in an implication and the Craig interpolation property, which states that a Craig interpolant exists if the implication is valid. We then introduce uniform interpolants, a strong form of Craig interpolants which do not depend on the right hand side of the implication and which can be regarded as a formalization of forgetting some propositional atoms in a formula. We then link the computation of Craig interpolants to the computation of explicit definitions and the Craig interpolation property to the (projective) Beth definability property. While conceptually, and in applications, definability and interpolation are equally important, because of their mutual reducibility it suffices to discuss one of the two notions in detail. We chose interpolation and next discuss generalizations of Craig interpolation to structured inputs containing more than two formulas and an application to the uniqueness of theory decompositions. Then we present, in the core of this chapter, four ways of computing Craig (and sometimes uniform) interpolants. We start by presenting a simple, but conceptually important, construction that introduces quantified propositional formulas, in which interpolants trivially exist, and then applies quantifier eliminations to obtain propositional logic interpolants. Next we show how interpolants are obtained from formulas in disjunctive normal form by dropping atoms. Our focus is, however, on the introduction of two slightly more sophisticated and also practical methods of computing interpolants, namely their construction from resolution and tableau refutations. In both cases we give rather detailed expositions, assuming no prior understanding of resolution or tableau techniques. We close by discussing the size of interpolants and linking it to open questions in circuit complexity, again assuming hardly any prior knowledge of complexity theory.

\section{Preliminaries}\label{sec:preliminaries}
\emph{Propositional formulas} are defined as usual by the grammar
\[
\varphi,\psi \quad := \quad p \quad | \quad \top \quad | \quad \bot \quad |\quad  \neg \varphi \quad | \quad \varphi \wedge \psi  \quad | \quad \varphi \vee \psi,
\]
where $p$ ranges over a countably infinite set of \emph{propositional
variables} or \emph{atoms}. We set
$\varphi \rightarrow \psi := \neg \varphi \vee \psi$ and
$\varphi \leftrightarrow \psi := (\varphi \rightarrow \psi) \land
(\psi \rightarrow \varphi)$.

A \emph{model} $v$ is a mapping from the set of atoms to $\{0,1\}$. The
\emph{truth value} $v(\varphi)\in \{0,1\}$ of a formula $\varphi$ under $v$ is
defined as usual by induction using truth tables. We set $v\models \varphi$
and say that $\varphi$ \emph{is satisfied in $v$} if $v(\varphi)=1$. A formula
$\varphi$ is \emph{valid}, in symbols $\models \varphi$, if $v\models \varphi$
for all models $v$, and $\varphi$ is \emph{satisfiable} if $v\models \varphi$
for some model $v$. We say that $\varphi$ \emph{entails} $\psi$, in symbols $\varphi \models \psi$, if $v\models \psi$ follows from $v\models
\varphi$, for all models $v$. Observe that $\varphi\models \psi$ iff $\models
\varphi\rightarrow \psi$. Formulas $\varphi$ and $\psi$ are \emph{logically
equivalent}, in symbols $\varphi \equiv \psi$, if $\varphi\models \psi$ and
$\psi\models \varphi$.

The set $\sub(\varphi)$ of \emph{subformulas} of a formula $\varphi$ is defined by induction by setting $\sub(p)=\{p\}$, $\sub(\top)=\{\top\}$,  $\sub(\bot)=\{\bot\}$, $\sub(\varphi\wedge \psi)=\{\varphi \wedge \psi\}\cup \sub(\varphi) \cup \sub(\psi)$,
$\sub(\varphi\vee \psi)=\{\varphi \vee \psi\}\cup \sub(\varphi) \cup \sub(\psi)$, and
$\sub(\neg\varphi)= \{\neg \varphi\} \cup \sub(\varphi)$. Define the size $|\varphi|$ of a propositional formula $\varphi$ as the
number of its distinct subformulas. This corresponds to representing formulas as directed acyclic graphs (dags) and is the standard measure used in circuit complexity theory.
We also call $|\varphi|$ the \emph{dag-size} of $\varphi$. 

In contrast, if one takes into account the number of different occurrences of subformulas, one obtains the \emph{tree-size} $s(\varphi)$ of $\varphi$, defined inductively by setting $s(p)=s(\top)=s(\bot)=1$, $s(\varphi\wedge \psi)=s(\varphi)+s(\psi)+1$, $s(\varphi\vee \psi)=s(\varphi)+s(\psi)+1$, and $s(\neg\varphi)=s(\varphi)+1$. Clearly, $s(\varphi)\geq |\varphi|$ and it is conjectured that there is a superpolynomial (even exponential) gap between dag-size and tree-size in the sense that there are no functions $e$ mapping every propositional formula $\varphi$ to a logically equivalent propositional formula $e(\varphi)$ and polynomial function $f$ (that is, a function of the form $f(n)=cn^{d}$ with $c,d$ natural numbers) such that $s(e(\varphi))\leq f(|\varphi|)$ for all $\varphi$.
However, this conjecture remains open~\cite{DBLP:journals/siamcomp/Rossman18}.

Central for interpolation is the notion of a signature: A \emph{signature} $\sigma$ is a finite set of atoms. The signature
$\sig(\varphi)$ of a formula $\varphi$ is the set of atoms that occur in
it. In other words, the signature of $\varphi$ are the non-logical symbols
that occur in $\varphi$, whereas logical symbols such as $\wedge$, $\vee$,
$\neg$, $\bot$ and $\top$ are not part of the signature of a formula.
If $\sig(\varphi) \subseteq \sigma$, we call $\varphi$ a \name{$\sigma$-formula}.
Given two formulas $\varphi$ and $\psi$, we call $\psi$ a \emph{conservative extension of $\varphi$} if
\red{$\psi \models \varphi$, $\sig(\varphi)\subseteq \sig(\psi)$, and for every $\sig(\varphi)$-formula $\chi$, $\psi\models\chi$ implies $\varphi\models\chi$.} %

Our methods for computing interpolants rely on different normal forms which we introduce next.
A \emph{literal} $\ell$ is either an atom or the negation $\neg p$ of an atom
$p$.
The \defname{complement} of a literal~$\ell$, written as $\complit{\ell}$,
is $\lnot p$ if $\ell$ is $p$, and $p$ if $\ell$ is $\lnot p$.
A disjunction of literals $C=\ell_{1}\vee \cdots \vee \ell_{n}$ is called a \emph{clause}, often written as $C=\{\ell_{1},\ldots,\ell_{n}\}$ and treated as set. As usual, the empty disjunction is identified with $\bot$. A conjunction of literals $C=\ell_{1} \wedge \cdots \wedge \ell_{n}$ is called a \emph{conjunctive clause}. The empty conjunction is identified with $\top$.

A formula $\varphi$ is in \emph{conjunctive normal form (CNF)} if it is a conjunction of clauses; it is in \emph{disjunctive normal form (DNF)} if it is a disjunction of conjunctive clauses. Every formula is logically equivalent to a formula in CNF and DNF.
For convenience, we may sometimes identify formulas in CNF/DNF with the corresponding sets of (conjunctive) clauses.

A formula $\varphi$ is in \emph{negation normal form (NNF)} if it is built
from literals, truth value constants, conjunction and disjunction. Formally, NNF
formulas are defined by the grammar
\[\varphi,\psi \quad := \quad p \quad | \quad \lnot p \quad | \quad \true
\quad | \quad \false \quad | \quad \varphi \wedge \psi \quad | \quad \varphi
\vee \psi,\]
where $p$ ranges over the set of propositional atoms.
CNF and DNF formulas are both special cases of NNF formulas.
While any propositional formula can be converted in linear time to an
equivalent NNF formula with an at most linear size increase,
the smallest CNF
and DNF formulas that are equivalent to a given formula may be exponentially
larger.\footnote{Each formula in DNF equivalent to $\bigwedge_{1\leq i \leq n}(p_{i} \vee q_{i})$ has at least $2^{n}$ conjunctive clauses. Each formula in CNF equivalent to $\bigvee_{1\leq i \leq n}(p_{i} \wedge q_{i})$ has at least $2^{n}$ clauses.}
One can, however, construct in polynomial time for every propositional formula $\varphi$ a formula $\varphi'$ in CNF that is a conservative extension of $\varphi$, using structural transformations~\cite{Baaz2001} such as the Tseitin encoding~\cite{Tseitin1968}.

\section{Craig Interpolants}
In this section we give the main definitions of this chapter.
\begin{definition}[Craig Interpolant]\label{def:ci}
    Let $\varphi,\psi$ be formulas. Then a formula $\chi$ is called a
    \emph{Craig interpolant} for $\varphi,\psi$ if
\begin{itemize}
	\item $\varphi \models \chi$ and $\chi \models \psi$; and
	\item $\sig(\chi) \subseteq \sig(\varphi) \cap \sig(\psi)$.
\end{itemize}
\end{definition}
Clearly, if a Craig interpolant for $\varphi,\psi$ exists, then $\varphi\models \psi$, by transitivity of the entailment relation $\models$. The converse direction, stating that a Craig interpolant for $\varphi,\psi$ exists whenever $\varphi\models \psi$, is called the \emph{Craig interpolation property (CIP)}. It holds for propositional logic.
\begin{theorem}[Craig Interpolation Property (CIP)]
  \label{thm-cip}
In propositional logic, if $\varphi\models \psi$, then there exists a Craig interpolant for $\varphi,\psi$.
\end{theorem}

We will present and discuss several proofs of this result in
Sections~\ref{sec:cip-by-quantifier-elimination}--\ref{sec-tableaux}. Here we
continue with illustrating the definition by examples, basic observations, and
variants of the definition.

\begin{example}
  \label{ex-cip}
  \ \par

  \subexample{ex-cip-basic} Observe that $p\wedge q_{1} \models q_{2}
  \rightarrow p$. By CIP a Craig interpolant for $p\wedge q_{1}$, $q_{2}
  \rightarrow p$ exists. Clearly $p$ is such a Craig interpolant. Moreover, up
  to logical equivalence, $p$ is the only Craig interpolant for $p\wedge
  q_{1}$, $q_{2} \rightarrow p$. Indeed, up to logical equivalence, the only
  other candidates are $\bot$, $\top$, and $\neg p$, and none of these is a
  Craig interpolant for $p\wedge q_{1}$, $q_{2} \rightarrow p$.

  \subexample{ex-cip-true-false} We have $p\wedge \neg p \models q$ and $\bot$
  is a Craig interpolant for $p\wedge \neg p$, $q$. Up to logical equivalence,
  $\bot$ is the only Craig interpolant. This example shows that without
  admitting at least one of $\top$ or $\bot \equiv \lnot \top$ as a formula not using any atoms,
  Craig interpolation would fail for formulas in disjoint signatures.

  \subexample{ex-cip-largespace} Consider atoms $p_{1},\ldots,p_{n}$. Then
  $\bigwedge_{1\leq i \leq n}(p_{i}\wedge \neg p_{i}) \models\bigvee _{1\leq i
    \leq n}(p_{i}\vee \neg p_{i})$ and any formula using only atoms from
  $p_{1},\ldots,p_{n}$ is a Craig interpolant.

  \subexample{ex-cip-many} Let $\varphi = \fp \land \fq \land \fr$ and $\psi =
  \fs \rightarrow (\fp \lor \fq)$. Then $\varphi \entails \psi$ and the Craig
  interpolants for $\varphi, \psi$ are, up to logical equivalence, the
  formulas $\fp \land \fq$, $\fp$, $\fq$, and $\fp \lor \fq$.

  \subexample{ex-cip-independent}
    If $\varphi$ is a propositional formula
  that is semantically independent from some atoms $\sigma \subseteq
  \sig(\varphi)$, i.e., $\varphi$ is equivalent to a propositional formula
  $\psi$ without atoms in $\sigma$, then Craig interpolation can be applied to
  find such a $\psi$. Specifically, any Craig interpolant for $\varphi,
  \varphi'$, where $\varphi'$ is the formula obtained from $\varphi$ by
  systematically replacing all atoms in $\sigma$ with fresh atoms, provides
  such a formula~$\psi$. For example, let
  \[\varphi = (\fp \rightarrow (\fq \land \fr)) \land
  (\fp \lor \fq) \land (\fq \rightarrow \fr).\] Let $\sigma = \{\fp\}$ and
  $\varphi' = (\fp' \rightarrow (\fq \land \fr)) \land (\fp' \lor \fq) \land
  (\fq \rightarrow \fr)$. We obtain $\psi = \fq \land \fr$, which is
  equivalent to $\varphi$, as Craig interpolant for $\varphi, \varphi'$.
  \lipicsEnd
\end{example}

Example~\ref{ex-cip-largespace} shows that the space of Craig interpolants for a fixed pair of formulas can be large. It has, however, a very transparent structure:
\begin{theorem}[Closure of Craig Interpolants under Conjunction and Disjunction]
    Let $\varphi$ and $\psi$ be formulas and $C(\varphi,\psi)$ be the set of
    Craig interpolants for $\varphi$, $\psi$. Then $C(\varphi,\psi)$ is closed under conjunctions and disjunctions: if $\chi_{1},\chi_{2}\in C(\varphi,\psi)$, then $\chi_{1} \wedge \chi_{2}\in C(\varphi,\psi)$ and $\chi_{1} \vee \chi_{2}\in C(\varphi,\psi)$.
\end{theorem}
As $C(\varphi,\psi)$ contains only finitely many formulas up to logical equivalence, it contains a logically strongest formula (that entails all others), given by the conjunction of all formulas in $C(\varphi,\psi)$, and a logically weakest formula (that is entailed by all others), given by the disjunction of all formulas in $C(\varphi,\psi)$.

In many applications and proofs, a different, but equivalent, view of Craig interpolants and CIP is useful.
\begin{definition}[Craig Separator]\label{def:separator} Let $\varphi$ and $\psi$ be formulas.
Then a formula $\chi$ is a \emph{Craig separator} for $\varphi, \psi$
if
\begin{itemize}
    \item $\varphi\models \chi$ and $\chi \wedge \psi\models \bot$; and
    \item $\sig(\chi) \subseteq \sig(\varphi) \cap \sig(\psi)$.
\end{itemize}
\end{definition}
Clearly, $\chi$ is a Craig separator for $\varphi$, $\psi$ iff $\chi$ is a Craig interpolant for $\varphi,\neg\psi$. The CIP is then equivalent to the statement
that for any formulas $\varphi,\psi$, if $\varphi \wedge \psi$ is not satisfiable, then there exists a Craig separator for $\varphi$ and $\psi$.

In the literature on verification, a Craig separator is often termed \name{reverse
  Craig interpolant} or even just \name{Craig interpolant}.

Craig interpolation can be strengthened to take also the \name{polarity} of
atom occurrences into account, which is often called \name{Craig-Lyndon}
interpolation. Let $\sigposneg(\varphi)$ denote the set of all pairs $\langle
p, \mathit{pol}\rangle$ such that atom $p$ occurs in formula $\varphi$ with
polarity $\mathit{pol} \in \{+,-\}$, where the polarity of an occurrence is
$+$ (respectively, $-$) if it is in the scope of an even (respectively, odd) number of negation symbols.
We can then define the notion of \name{Craig-Lyndon interpolant} as follows.

\begin{definition}[Craig-Lyndon Interpolant]\label{def:cli}
    Let $\varphi,\psi$ be formulas. Then a formula $\chi$ is called a
    \emph{Craig-Lyndon interpolant} for $\varphi,\psi$ if
\begin{itemize}
	\item $\varphi \models \chi$ and $\chi \models \psi$; and
	\item $\sigposneg(\chi) \subseteq \sigposneg(\varphi) \cap
          \sigposneg(\psi)$.
\end{itemize}
\end{definition}

\begin{example}
  \label{ex-lip} Let $\varphi = (p \imp q) \land (r \imp (p \lor q)) \land t$
  and let $\psi = (p \imp (q \land t)) \land ((q \land s) \imp t)$. Atom~$p$
  occurs in~$\varphi$ with both polarities but in $\psi$ only negatively.
  Atom~$q$ occurs in $\psi$ with both polarities but in $\varphi$ only
  positively. Atom~$t$ occurs just positively, in both $\varphi$ and~$\psi$.
  Hence, in a Craig-Lyndon interpolant for $\varphi, \psi$ the atom $p$ is
  allowed only negatively, and $q, t$ are allowed only positively. Atoms $r,
  s$ are not allowed at all, since each of them occurs just in one of
  $\varphi$ or $\psi$. Indeed, $\chi = (p \imp q) \land t$ is a Craig-Lyndon
  interpolant for $\varphi, \psi$. The formula $\chi' = (p \imp (q \land t))
  \land (q \imp t)$ is another Craig interpolant for $\varphi,\psi$, but it is
  no Craig-Lyndon interpolant as it has a negative occurrence of $q$.
\lipicsEnd
\end{example}

A Craig-Lyndon interpolant is sometimes also called a \name{Lyndon
  interpolant} or a Craig interpolant with the \name{Lyndon property}. In
propositional logic, the following strengthening of
Theorem~\ref{thm-cip} holds.

\begin{theorem}[Craig-Lyndon Interpolation Property (LIP)]
  \label{thm-cip-lyndon}
In propositional logic, if $\varphi\models \psi$, then there exists a
Craig-Lyndon interpolant for $\varphi,\psi$.
\end{theorem}

Since a Craig-Lyndon interpolant is also a Craig interpolant, the LIP
(Theorem~\ref{thm-cip-lyndon}) implies the CIP (Theorem~\ref{thm-cip}). In
fact, some of the proofs in
Sections~\ref{sec:cip-by-quantifier-elimination}--\ref{sec-tableaux} actually
show the LIP. The notion of \name{Craig separator} also has an analog where polarity
is considered.

\begin{definition}[Craig-Lyndon Separator]
   Let $\varphi$ and $\psi$ be formulas.
Then a formula $\chi$ is a \emph{Craig-Lyndon separator} for $\varphi, \psi$
if
\begin{itemize}
    \item $\varphi\models \chi$ and $\chi \wedge \psi\models \bot$; and
    \item $\sigposneg(\chi) \subseteq \sigposneg(\varphi) \cap
      \sigposneg(\lnot \psi)$.
\end{itemize}
\end{definition}
As before, a formula~$\chi$ is a Craig-Lyndon separator for $\varphi$, $\psi$ iff $\chi$ is a Craig-Lyndon interpolant for $\varphi,\neg\psi$.

\section{Uniform Interpolants}
\label{section:uniform}
In this section we introduce a special form of Craig interpolant called a \name{uniform} interpolant. While Craig interpolants are defined in terms of a pair of input formulas $\varphi,\psi$ with $\varphi\models\psi$, uniform interpolants are defined in terms of a single formula $\varphi$ and a signature $\sigma\subseteq \sig(\varphi)$.
\begin{definition}[Uniform Interpolant]\label{def:uniform-interpolant}
Let $\varphi$ be a propositional formula and $\sigma\subseteq \sig(\varphi)$. Then a propositional formula $\chi$ is called a
\emph{uniform $\sigma$-interpolant for $\varphi$} if
\begin{itemize}
    \item $\varphi \models \chi$;
    \item \red{$\sig(\chi)\subseteq \sigma$}; and
    \item for every propositional formula $\psi$, if $\varphi \models \psi$ and \red{$\sig(\psi) \cap \sig(\varphi) \subseteq \sigma$}, then $\chi\models \psi$.\footnote{Uniform interpolants are not always defined in exactly the same way. For instance, instead of starting from the signature $\sigma$ of the uniform interpolant, one can start from the signature $\sig(\varphi)\setminus\sigma$ that is `forgotten', or the condition $\sig(\psi) \cap \sig(\varphi) \subseteq \sigma$ is replaced by the condition $\sig(\psi) \subseteq \sigma$. In the context of propositional logic these different definitions are equivalent.}
\end{itemize}
\end{definition}
Observe that, given $\varphi$ and $\sigma\subseteq \sig(\varphi)$, a uniform $\sigma$-interpolant for $\varphi$ is a Craig interpolant for all pairs $\varphi,\psi$ such that $\varphi\models \psi$ and $\sigma=\sig(\psi)\cap \sig(\varphi)$. In contrast to Craig interpolants, uniform interpolants are uniquely determined (up to logical equivalence) by $\varphi$ and $\sigma$. \red{In what follows, we therefore refer to \emph{the} uniform $\sigma$-interpolant for a formula $\varphi$.}

\begin{example}
  \label{ex-uniform}
  We consider the uniform $\sigma$-interpolants for the formulas $\varphi$ in the entailments $\varphi\models \psi$ in Example~\ref{ex-cip}, \red{where $\sigma=\sig(\varphi)\cap\sig(\psi)$}. The uniform $\{p\}$-interpolant for $p\wedge q_{1}$ is given by $p$ and the uniform $\emptyset$-interpolant for $p\wedge \neg p$ is given by $\bot$. Similarly, the uniform \red{$\{p_{1},\ldots,p_{n}\}$-interpolant} for $\bigwedge_{1\leq i \leq n}(p_{i}\wedge \neg p_{i})$ is given by $\bot$. Finally, the uniform
  \red{$\{p,q\}$-interpolant} for $p \wedge q \wedge r$ is $p\wedge q$.
\lipicsEnd
\end{example}
Note that in Example~\ref{ex-uniform}, the uniform $\sigma$-interpolant is always the logically strongest Craig interpolant for $\varphi,\psi$. This is no accident.
To show this in general, assume that $\chi$ is the logically strongest Craig interpolant for $\varphi,\psi$. Let \red{$\sigma=\sig(\varphi) \cap \sig(\psi)$}
and let $\chi'$ be the uniform $\sigma$-interpolant for $\varphi$. We show that
$\chi$ and $\chi'$ are logically equivalent. Clearly $\chi'\models \chi$ since $\varphi\models \chi$ and \red{$\sig(\chi)\subseteq \sigma$.}
We also have $\chi\models \chi'$ since $\chi'$ is a Craig interpolant for $\varphi,\psi$ and $\chi$ is the logically strongest one.

Sometimes, a uniform interpolant as in Definition~\ref{def:uniform-interpolant} is also called \emph{right-uniform interpolant}, indicating that it is a Craig interpolant for an implication $\varphi\models\psi$ that is independent of the formula \emph{on the right}. This notion can then be contrasted with that of a \emph{left-uniform interpolant}, which is a Craig interpolant that is independent of the formula on the left (it is a Craig interpolant also for any other entailment $\varphi'\models\psi$):
\begin{definition}[Left-Uniform Interpolant]\label{def:left-uniform-interpolant}
Let $\varphi$ be a propositional formula and $\sigma\subseteq \sig(\varphi)$. Then a propositional formula $\chi$ is called a
\emph{left-uniform $\sigma$-interpolant for $\varphi$} if
\begin{itemize}
    \item $\chi\models\varphi$; 
    \item
    \red{$\sig(\chi)\subseteq \sigma$}; and
    \item for every propositional formula $\psi$, if $\psi\models\varphi$ and \red{$\sig(\psi) \cap \sig(\varphi) \subseteq \sigma$}, then $\psi\models\chi$.
\end{itemize}
\end{definition}
\red{For logics with classical negation}, right-uniform interpolation can easily be reduced to left-uniform interpolation and vice versa: $\chi$ is a right-uniform $\sigma$-interpolant for $\psi$ iff $\neg\chi$ is a left-uniform $\sigma$-interpolant for $\neg\psi$. For the rest of this chapter, we therefore focus on right-uniform interpolants and call them simply uniform interpolants.

The following result is shown in different ways in Sections~\ref{sec:cip-by-quantifier-elimination}--\ref{sec:cip-by-resolution}.
\begin{theorem}[Uniform Interpolation Property (UIP)]\label{thm:uniform}
In propositional logic, if $\varphi$ is a propositional formula and $\sigma\subseteq \sig(\varphi)$, then a uniform $\sigma$-interpolant for $\varphi$ exists.
\end{theorem}
Note that any logic with the uniform interpolation property automatically satisfies the CIP.

\section{Beth Definability}
In this section, we show that Craig interpolants are closely related to definitions and can be used to compute an explicit definition of an atom that is implicitly defined. We first formalize the notions of explicit and implicit definitions.

\begin{definition}[Explicit Definability] Let $\varphi$ be a propositional
  formula, $\sigma$ be a set of atoms, and $p\not\in\sigma$. Then a propositional formula $\psi$ is a \name{$\sigma$-definition of $p$ under $\varphi$}
  if $\sig(\psi)\subseteq \sigma$ and 
	$\varphi\models p \leftrightarrow \psi$.
    $p$ is called \emph{explicitly $\sigma$-definable under $\varphi$} if there exists a $\sigma$-definition of $p$ under $\varphi$.
\end{definition}
The condition $p\not\in\sigma$ reflects the intuition that definitions should not be circular. Also note that we are not stipulating a definition of $p$ but ask whether $p$ is logically equivalent to some $\psi$ under the assumption $\varphi$.

Alternatively, we may approach definability from a semantic point of view: The
atom $p$ is \name{implicitly $\sigma$-definable under}
$\varphi$ if, whenever two models that satisfy $\varphi$ agree on the values
of all atoms in $\sigma$, then they also agree on the value of $p$.
Intuitively, the interpretation of the atoms in $\sigma$ fully determines
the interpretation of $p$. The formal definition is as follows.

\begin{definition}[Implicit Definability] Let $\varphi$ be a propositional formula, $\sigma$ be a set of atoms, and $p\not\in\sigma$. Then $p$ is \emph{implicitly $\sigma$-definable under $\varphi$} if for any models $v_{1},v_{2}$ satisfying $\varphi$,
$v_{1}(q)=v_{2}(q)$ for all $q\in \sigma$ implies $v_{1}(p)=v_{2}(p)$.
\end{definition}
Clearly explicit definability implies implicit definability. The converse is called \emph{projective Beth definability property (BDP)} and does not hold for all logics.
It does hold, however, for classical propositional logic.
\begin{theorem}[Projective Beth Definability Property (BDP)]
  \label{thm-beth}
  Propositional logic has the projective Beth definability property.
\end{theorem}
\begin{proof}
The proof is by reduction to the Craig interpolation property.
Assume $p$ is implicitly $\sigma$-definable under $\varphi$. Let $\varphi'$ be
obtained from $\varphi$ by systematically replacing all atoms $q$ not in
$\sigma$ by associated fresh atoms $q'$. Then implicit definability implies
	\[
	\varphi\wedge \varphi' \models p\leftrightarrow p'
	\]
	Hence
	\[
	\varphi \wedge p \models \varphi' \rightarrow p'
	\]
Let $\chi$ be any Craig interpolant for $\varphi \wedge p,\varphi' \rightarrow p'$. We show that $\chi$ is a $\sigma$-definition of $p$ under $\varphi$. By definition of Craig interpolants, $\sig(\chi)\subseteq \sigma$. From $\varphi \wedge p \models \chi$ we obtain $\varphi \models p\rightarrow \chi$. Conversely,
from $\chi\models \varphi'\rightarrow p'$ we obtain $\varphi'\models \chi \rightarrow p'$. Then, since
$\sig(\chi)\subseteq \sigma$, we obtain $\varphi\models \chi \rightarrow p$, as required.
\end{proof}
Observe that the proof of Theorem~\ref{thm-beth} shows how one can construct \red{in polynomial time} explicit definitions from Craig interpolants: given $\varphi$, a signature $\sigma$, and $p\not\in \sigma$, a $\sigma$-definition of $p$ under $\varphi$ is obtained by taking any Craig interpolant
for propositional formulas $\chi,\chi'$ computed from $\varphi$, $\sigma$, and $p$ in linear time. 

The special case of the projective Beth definability property where $\sigma$ is not
a parameter but defined as $\sigma=\sig(\varphi)\setminus \{p\}$ is known as the \name{Beth definability property}. Note also that the notions of explicit and implicit definability, 
and the projective Beth definability property can be easily generalized to definitions 
of compound propositional formulas in place of atoms $p$.

\begin{example}
  \ 
  
  \subexample{ex-beth-verysimple} As a very simple first example for Beth definability consider
  $\varphi = p \equi q$ and $\sigma = \{q\}$. Clearly $p$ is
  $\sigma$-definable under $\varphi$. The $\sigma$-definition $\fq$ is a Craig
  interpolant for $(\varphi \land \fp)$,$(\varphi(\fp/\fp') \imp
  \fp')$. The entailment underlying the Craig interpolation is
  $(p \equi q) \land \fp \entails (p' \equi q) \imp \fp'$, which
  can be equivalently expressed as $p \land q \entails
  p' \lor q$.

  \subexample{ex-beth-one} Let $\varphi = (\fp \imp (\fq \land \fr)) \land
  ((\fq \land \fs) \imp \fp) \land (\fr \imp \fs)$ and $\sigma = \{\fq,
  \fr\}$. Since $\varphi \entails \fp \equi (\fq \land \fr)$ atom $\fp$ is
  $\sigma$-definable under $\varphi$. The $\sigma$-definition $\fq \land \fr$
  is a Craig interpolant for $(\varphi \land
  \fp)$,$(\varphi(\fp/\fp',\fs/\fs') \imp \fp')$.

  \subexample{ex-beth-two} Let $\varphi = (\fp \imp (\fq \land \fr)) \land
  ((\fq \lor \fr) \imp \fp)$ and $\sigma = \{\fq, \fr\}$. Then $\fp$ is
  $\sigma$-definable under $\varphi$, actually with several
  $\sigma$-definitions: $\varphi \entails \fp \equi (\fq \land \fr)$, $\varphi
  \entails \fp \equi \fq$, $\varphi \entails \fp \equi \fr$, and $\varphi
  \entails \fp \equi (\fq \lor \fr)$. Each of these four definitions is a
  Craig interpolant for $(\varphi \land \fp)$,$(\varphi(\fp/\fp') \imp
  \fp')$.

  \subexample{ex-beth-three} Let $\varphi = \fp$ and $\sigma = \{\}$. Then
  $\fp$ is $\sigma$-definable under $\varphi$, since $\fp \entails \fp \equi
  \top$. The $\sigma$-definition $\top$ is a Craig interpolant of $(\varphi
  \land \fp)$,$(\varphi(\fp/\fp') \imp \fp')$, i.e., of $(\fp \land
  \fp)$,$(\fp' \imp \fp')$.
  \lipicsEnd
\end{example}

We also give a polynomial time reduction of the computation of Craig interpolants to the computation of explicit definitions.

\begin{theorem}[Reduction of Craig Interpolation to Explicit Definitions]\label{thm:converse}
Assume $\models \varphi \rightarrow \psi$. Then the following conditions are equivalent
for any formula $\chi$:
\begin{itemize}
	\item $\chi$ is a Craig interpolant for $\varphi,\psi$;
	\item $\chi$ is an explicit $\sig(\varphi)\cap \sig(\psi)$-definition of $\psi$ under $\psi\rightarrow \varphi$.
\end{itemize}
\end{theorem}
Because of the mutual polynomial time reduction between Craig interpolant computation and explicit definition computation shown in this section, in what follows we focus on Craig interpolation without always mentioning again the consequences for explicit definitions.

\section{Generalisations and an Application to Theory Decomposition}
Craig interpolation has been generalised in many different ways. In particular, in applications of Craig interpolation to program verification various inductive generalisations have been introduced, including inductive sequences of interpolants 
\cite{DBLP:conf/popl/HenzingerJMM04,DBLP:conf/cav/McMillan06},
tree interpolants \cite{DBLP:conf/popl/HeizmannHP10,DBLP:conf/lpar/BlancGKK13},
and disjunctive interpolation \cite{DBLP:conf/cav/RummerHK13,DBLP:journals/fmsd/RummerHK15}, in increasing order of generality. We refer the reader to \refchapter{chapter:automated} for further discussion. In all these generalisations, one considers multiple formulas (for instance, as labels of nodes in a tree), and the interpolants capture the interaction between these formulas in their shared signature.

Here we focus on a generalisation observed by Craig already \cite[Lemma~2]{craig:uses}. 
\begin{theorem}\label{thm:parallel1}  In propositional logic, if $\varphi_1 \land \ldots \land \varphi_k \entails \varphi_{k+1} \lor
    \ldots \lor \varphi_l$, then there are formulas $\chi_{1},\ldots,\chi_{l}$ with
    \begin{itemize}
    \item $\varphi_i \entails \chi_i$ for $1 \leq i \leq k$
    \item $\chi_i \entails \varphi_i$ for $k+1 \leq i \leq l$
    \item $\sig(\chi_i) \subseteq \sig(\varphi_i) \cap
      \bigcup_{j \in \{1, \ldots, l\}, j \neq i} \sig(\varphi_j)$
      for $1 \leq i \leq l$
    \item $\chi_1 \land \ldots \land \chi_k \entails
      \chi_{k+1} \lor \ldots \lor \chi_l$
\end{itemize}
\end{theorem}
In applications to program verification, these interpolants have been studied for $l=k+1$ and $\varphi_{k+1}=\bot$~\cite[Sect.~5]{mcmillan:symmetric:2005}, where they are called \name{symmetric interpolants}. We focus on a brief discussion of another special case of Theorem~\ref{thm:parallel1} called \emph{parallel interpolation}.
In what follows, $T$ denotes a finite set of propositional formulas which, intuitively, axiomatises some theory or knowledge base of interest. We often identify $T$ with the conjunction $\bigwedge_{\chi\in T}\chi$; for example, $T\models \varphi$ stands for $\bigwedge_{\chi\in T}\chi\models \varphi$. Now, parallel interpolation is the special case in which $l=k+1$ and $\sig(\varphi_{i})\cap \sig(\varphi_{j})=\emptyset$ for $1\leq i<j\leq k$. So, given $T=\{\varphi_{1},\ldots,\varphi_{n}\}$ with $\sig(\varphi_{i})\cap \sig(\varphi_{j})=\emptyset$ for $1\leq i<j\leq n$
and a formula $\psi$ with $T\models \psi$, there are formulas $\chi_1,\ldots,\chi_{n}$, called \emph{parallel interpolants}, with
   	\begin{itemize}
   		\item $\varphi_{i}\models \chi_{i}$ for $1\leq i \leq n$;
   		\item $\sig(\chi_{i}) \subseteq \sig(\varphi_{i})\cap \sig(\psi)$ for $1\leq i \leq n$; and
   		\item $\{\chi_{1},\ldots,\chi_{n}\} \models \psi$.
    	\end{itemize}
Intuitively, the parallel interpolants $\chi_{i}$ capture, in the shared atoms of $\varphi_{i}$ and $\psi$, the contribution of $\varphi_{i}$ to the derivation of $\psi$ from $T$. Although fairly straightforward to derive from Craig interpolation, parallel interpolation has some unexpected applications. Here we use it to show that finest decompositions of theories into axioms with mutually disjoint signatures are uniquely determined (up to logical equivalence). In detail, let $T$ be a set of formulas and let $\sigma_{1},\ldots,\sigma_{n}$ be a partition of $\sig(T)$. Then $\sigma_{1},\ldots,\sigma_{n}$ is called a \emph{splitting} of $T$ if there are formulas $\chi_{1},\ldots,\chi_{n}$ such that $\sig(\chi_{i})\subseteq \sigma_{i}$ for $1\leq i \leq n$ and $\{\chi_{1},\ldots,\chi_{n}\}$ axiomatises $T$,
that is to say, $\bigwedge_{1\leq i \leq n}\chi_{i}$ and $\bigwedge_{\chi \in T}\chi$ are logically equivalent. Splittings formalise the idea of decomposing a theory $T$ into axioms that speak about mutually disjoint subject matters. 
We say that a partition $\sigma_{1},\ldots,\sigma_{n}$ of $\sig(T)$ is at least as fine as partition $\sigma_{1}',\ldots,\sigma_{m}'$ $\subseteq\sig(T)$ if for all $\sigma_{i}'$ there exists $\sigma_{j}$ with $\sigma_{j}\subseteq \sigma_{i}'$.

\begin{theorem}
    Every finite set $T$ of propositional formulas has a unique finest splitting.
\end{theorem}
\begin{proof}
Assume splittings $\sigma_{1},\ldots,\sigma_{n}$ and $\sigma_{1}',\ldots,\sigma_{m}'$ of $T$ are given. We show that the set of 
non-empty $\sigma_{i}\cap \sigma_{j}'$, with $1\leq i \leq n$ and $1\leq j\leq m$ is again a splitting of $T$. The claim then follows directly by assuming that $\sigma_{1},\ldots,\sigma_{n}$ and $\sigma_{1}',\ldots,\sigma_{m}'$ are finest splittings of $T$. Let $\varphi_{1},\ldots,\varphi_{n}$ and $\psi_{1},\ldots,\psi_{m}$ be axiomatisations of $T$ witnessing that $\sigma_{1},\ldots,\sigma_{n}$ and $\sigma_{1}',\ldots,\sigma_{m}'$ are splittings of $T$, respectively.
Then take the parallel interpolants $\chi_{ij}$ for the entailments $\{\varphi_{1},\ldots,\varphi_{n}\}\models \psi_{j}$ with $\varphi_{i}\models \chi_{ij}$, for $1\leq i \leq n$ and $1\leq j\leq m$. Then $\sig(\chi_{ij})\subseteq \sigma_{i}\cap \sigma_{j}'$ and the $\chi_{ij}$ axiomatise $T$. So they witness that the non-empty $\sigma_{i}\cap \sigma_{j}'$ form a splitting of $T$. 
       \qedhere
\end{proof}
For further results on parallel interpolation and its application to theory decomposition in philosophy and knowledge representation we refer the reader to~\cite{DBLP:journals/jsyml/KourousiasM07,DBLP:conf/kr/KonevLPW10,DBLP:journals/synthese/Parikh11}.

\section{Interpolants via Quantifier Elimination}
\label{sec:cip-by-quantifier-elimination}
 Having discussed various generalizations and applications of interpolation, in the next few sections, we discuss different proofs of the Craig interpolation property (CIP) and methods for computing interpolants.
 We start with the easiest proof, which shows the stronger result that propositional logic has the uniform interpolation property (\Cref{thm:uniform}). CIP then follows. While the proof constructs uniform interpolants (and so Craig interpolants), it does not show how to construct interpolants efficiently. We examine efficient methods in Sections~\ref{sec:cip-by-dnf}--\ref{sec-tableaux}.
 
 The proof has two steps: (i) we show that uniform interpolants %
 always exist in the language obtained from propositional logic by adding quantifiers over atoms; and (ii) we show that these quantifiers can be always eliminated in the sense that one can always construct a logically equivalent propositional formula that does not use quantifiers. Overall, one then obtains that uniform interpolants %
 always exist.

We first extend propositional logic by adding quantifiers. \name{QBF (quantified Boolean formulas)} are an extension of propositional logic defined
according to the following grammar:
\[
\varphi,\psi \quad := \quad p \quad | \quad \top \quad | \quad \bot \quad | \quad  \neg \varphi \quad | \quad \varphi \wedge \psi  \quad | \quad \varphi \vee \psi \quad | \quad \exists p. \varphi
\]
where $p$ ranges over a countably infinite set of atoms. Models $v$ of QBF are defined in the same way as for propositional logic and we define $v\models \varphi$ inductively using the additional condition
    \begin{itemize}
    	\item $v\models \exists p.\varphi$ if there is a model $v'$ that coincides with $v$ for all atoms except possibly $p$ such that $v'\models \varphi$.
    \end{itemize}
Validity, satisfiability, and logical equivalence are now defined in the obvious way. The definition of the \emph{signature} $\sig(\varphi)$ of a QBF $\varphi$ extends inductively the definition for  propositional logic with $\sig(\exists p.\varphi) = \sig(\varphi)\setminus \{p\}$.

We are now in a position to give the mentioned proof of uniform interpolation in two steps.

\medskip
\noindent
\emph{QBF Uniform Interpolants.} Assume $\varphi$ and a signature $\sigma\subseteq \sig(\varphi)$ are  given. Let $p_{1},\ldots,p_{n}$ be  any ordering of the atoms in \red{$\sig(\varphi) \setminus \sigma$} and consider the QBF $\chi=\exists p_{1} \cdots \exists p_{n}.\varphi$.
Then
\begin{itemize}
    \item $\varphi\models \chi$ and $\sig(\chi) = \sigma$;
    \item for every propositional formula $\psi$, if $\varphi\models\psi$ and \red{$\sig(\psi) \cap \sig(\varphi) \subseteq \sigma$}, then $\chi\models\psi$.
\end{itemize}
and so $\chi$ satisfies the conditions for a  uniform $\sigma$-interpolant for $\varphi$, except that it is not a propositional formula.

\medskip
\noindent
\emph{Quantifier Elimination.} Let $\varphi=\exists p_{1} \cdots \exists p_{n}.\psi$ be a QBF with $\psi$ a propositional formula. Then $\varphi$ is logically equivalent to a propositional formula. Indeed, let $S_{n}=\{\bot,\top\}^{n}$ be the set of all sequences of $\bot$ and $\top$ of length $n$.
Then
\[
\bigvee_{t_{1},\ldots,t_{n}\in S_{n}}\psi[p_{1}/t_{1},\ldots,p_{n}/t_{n}]
\]
is logically equivalent to $\varphi$, where $\psi[p_{1}/t_{1},\ldots,p_{n}/t_{n}]$ is obtained from $\psi$ by replacing $p_{i}$ by $t_{i}$ for $1\leq i \leq n$. Using quantifier elimination, we can thus easily compute propositional uniform interpolants from QBF uniform interpolants.

Note that it can now also be shown by  induction that every QBF is logically equivalent to a propositional formula.

\begin{example}
  Let $\varphi = \exists \fq . (\fp \rightarrow \fq) \land (\fq \rightarrow
  \fr)$. By quantifier elimination $\varphi$ is equivalent to
  $((\fp \rightarrow \bot) \land (\bot \rightarrow \fr)) \lor
  ((\fp \rightarrow \top) \land (\top \rightarrow \fr))$, which simplifies
  to $\fp \rightarrow \fr$.
\lipicsEnd
\end{example}
We conclude this section with a few further observations.
Let $\forall p.\varphi:=\neg \exists p. \neg \varphi$. When constructing Craig interpolants for formulas $\varphi,\psi$, instead of $\exists p_{1}\cdots \exists
p_{n}.\varphi$ with $p_{1},\ldots,p_{n}$ any ordering of $\sig(\varphi)\setminus \sig(\psi)$, we could have also used $\forall q_{1}\cdots \forall
q_{n}.\psi$ with $q_{1},\ldots,q_{n}$ any ordering of $\sig(\psi)\setminus
\sig(\varphi)$.
The formula $\exists
p_{1}\cdots \exists p_{n}.\varphi$ is (up to equivalence) the logically strongest interpolant for $\varphi,\psi$ (it
entails all others) and the formula $\forall q_{1} \cdots \forall q_{n}.\psi$ is (up to equivalence) the
logically weakest interpolant for $\varphi,\psi$ (it is entailed by all others).

Quantifier elimination generalizes deciding
satisfiability, which can be seen as follows. For $\exists p_1 \ldots \exists p_n.\varphi $ where $\{p_1,
\ldots, p_n\} = \sig(\varphi)$ quantifier elimination yields a propositional
formula built just from logical symbols. Rewriting with equivalences $\lnot
\bot \equiv \top$, $\lnot \top \equiv \bot$, $\top \land \psi \equiv \psi$, $\bot \land \psi
\equiv \bot$, $\top \lor \psi \equiv \top$, and $\bot \lor \psi \equiv
\psi$ yields $\top$ if $\varphi$ is satisfiable and $\bot$ if $\varphi$ is
unsatisfiable.

\section{Interpolants via DNF}\label{sec:cip-by-dnf}
The uniform $\sigma$-interpolant for a formula $\varphi$ formalises a natural notion of forgetting the atoms in \red{$\sig(\varphi)\setminus \sigma$} from $\varphi$. This suggests that one can construct the uniform $\sigma$-interpolant by somehow dropping the atoms in \red{$\sig(\varphi)\setminus \sigma$} from $\varphi$ in a syntactic sense. This is clearly not the case in general.
For instance, for $\varphi=(p \vee q) \wedge \neg p$ and \red{$\sigma=\{p\}$}, we do not obtain
the uniform $\sigma$-interpolant for $\varphi$ by just dropping $q$: the result of dropping $q$ is logically equivalent to $p\wedge \neg p$ and $\varphi\not\models p\wedge \neg p$.
If, however, $\varphi$ is in DNF, then we indeed obtain the uniform $\sigma$-interpolant
by dropping the atoms in $\sig(\varphi)\setminus\sigma$.

Assume $\varphi$ is given in DNF, say
\[
\varphi = \bigvee_{1\leq i \leq n}\varphi_{i}, \quad \varphi_{i}=\bigwedge_{1\leq j \leq n_{i}}\ell_{ij}
\]
with $\ell_{ij}$ literals. Let $\sigma\subseteq \sig(\varphi)$. We may assume that no conjunctive clause $\varphi_{i}$ contains an atom and its negation. Then obtain $\varphi^{-\sigma}$ from $\varphi$ by dropping all literals $\ell_{ij}$ not in $\sigma$ from every $\varphi_{i}$.
\begin{lemma}
  \label{lem-dnf-uniform}
    Let $p_{1},\ldots,p_{n}$ be an ordering of \red{$\sig(\varphi)\setminus \sigma$}. Then $\varphi^{-\sigma}$ is logically equivalent to $\exists p_{1} \cdots \exists p_{n}. \varphi$.
\end{lemma}
\begin{proof}
  The lemma can be proven as follows with a syntactic argument.
  We can move the
  existential quantifiers in $\exists p_{1} \cdots \exists p_{n}. \varphi$
  inward, preserving equivalence, until all quantifications are of the form
  $\exists p_i. p_i$ or $\exists p_i. \lnot p_i$ (we assume that the
  conjunctive clauses of $\varphi$ contain no duplicate literals).
  We can do this due to the following equivalences: 1) $\exists p_1 \ldots \exists
  p_n . \varphi \lor \psi \equiv \exists p_1 \ldots \exists p_n . \varphi \lor
  \exists p_1 \ldots \exists p_n . \psi$; 2) if $p \notin \sig(\psi)$, then
  $\exists p . \varphi \land \psi \equiv (\exists p . \varphi) \land \psi$;
  and 3) if $p \notin \sig(\varphi)$, then $\exists p . \varphi \equiv \varphi$.

  Then we
  apply our quantifier elimination method from
  \Cref{sec:cip-by-quantifier-elimination} to these quantified literals. For
  positive as well as negative literals this yields $\top$, i.e., $\exists p_i.
  p_i \equiv p_i[p_i/\bot] \lor p_i[p_i/\top] \equiv \top$ and $\exists
  p_i. \lnot p_i \equiv \lnot p_i[p_i/\bot] \lor \lnot p_i[p_i/\top] \equiv \top$.
  By removing these redundant subformulas $\top$ from the conjunctive clauses,
  we obtain exactly $\varphi^{-\sigma}$.
\end{proof}

We can generalize this idea of computing uniform interpolants to the computation of Craig interpolants, for which we take into account
the logical strength of the right hand side of the entailment.
It is more convenient to look at Craig separators here (the construction of Craig interpolants would assume the 
right-hand side to be in CNF rather than DNF).
Assume that
\[
\varphi = \bigvee_{1\leq i \leq n}\varphi_{i}, \quad \varphi_{i}=\bigwedge_{1\leq j \leq n_{i}}\ell_{ij},
\quad \psi = \bigvee_{1\leq i' \leq m}\psi_{i'}, \quad \psi_{i'}=\bigwedge_{1\leq j' \leq m_{i'}}\ell_{i'j'}'
\]
and that $\varphi \wedge \psi$ is not satisfiable. We may assume that no
$\varphi_{i}$ nor $\psi_{i}$ contains an atom and its negation. We construct a
Craig separator as follows. Let $\sigma= \sig(\varphi) \cap \sig(\psi)$.
Because $\varphi \wedge \psi$ is not satisfiable, for any pair $i,i'$ we find
$j,j'$ s.t. $\ell_{ij}\wedge \ell_{i'j'}'$ unsatisfiable. Take any collection
$P$ of such pairs $\tup{\ell_{ij},\ell_{i'j'}'}$. Then $P$ can be regarded as
a \emph{proof} that $\varphi\wedge \psi$ is not satisfiable. Observe that each
pair in $P$ consists of an atom in $\sigma$ and its negation.
Now, let $\varphi_{P}$ be the DNF formula obtained from $\varphi$ by dropping
all literals $\ell_{i,j}$ that have no matching partner in $P$, i.e., that do
not occur as first component of some pair in $P$.
It is easy to check that $\varphi_P$ satisfies all properties of a Craig separator:
$\varphi\models \varphi_{P}$, $\sig(\varphi_{P})\subseteq \sigma$, and $\varphi_{P} \wedge \psi$ is unsatisfiable since the pairs in $P$ still show
that $\varphi_{P}\wedge \psi$ is unsatisfiable. Indeed, it is even a Craig-Lyndon separator: Any literal in $\varphi_{P}$ occurs as a literal $\ell_{ij}$ in $\varphi$, while its complement occurs as a literal $\ell_{i'j'}'$ in $\psi$. We obtain the following theorem, which is at the same time our first proof of \Cref{thm-cip-lyndon}, namely that propositional logic has the LIP.
\begin{theorem}
    $\varphi_{P}$ is a Craig-Lyndon separator for $\varphi,\psi$.
\end{theorem}

\begin{example}
  Let $\varphi = \fp \land \fq \land \fr$ and $\psi = \fs \rightarrow (\fp
  \lor \fq)$, as in Example~\ref{ex-cip-many}, and let \red{$\sigma = \sig(\varphi)
  \cap \sig(\psi) = \{\fp,\fq\}$}. Formula $\varphi$ is already in DNF, a
  single conjunctive clause. As $\varphi^{-\sigma}$ we obtain $\fp \land \fq$
  by dropping atom $\fr$ from this conjunctive clause. The formula $\fp \land
  \fq$ is the \red{unique uniform $\{\fp,\fq\}$-interpolant for $\varphi$}. Now consider
  $\psi' = \fs \land \lnot \fp \land \lnot \fq$, the negation of $\psi$ in
  DNF, also a single conjunctive clause. A collection of pairs $\tup{\ell_{ij},
  \ell_{i'j'}'}$ as described above is
    $\{(\ell_{11}, \ell_{12}')\} = \{(\fp, \lnot
  \fp)\}$. From this pair, we obtain %
  $\fp$ as Craig separator for $\varphi, \psi'$ (and as Craig interpolant
  for $\varphi,\psi$). Another suitable set of pairs is $\{(\ell_{12},
  \ell_{13}')\} = \{(\fq, \lnot \fq)\}$. Based on this we obtain $\fq$ as
  another Craig separator.
\lipicsEnd  
\end{example}

\section{Interpolants via Resolution}\label{sec:cip-by-resolution}

The quantifier elimination method described in \Cref{sec:cip-by-quantifier-elimination} gives an elegant proof of the CIP, and a simple method for computing Craig and uniform interpolants. A downside however is that the interpolants computed in this way are formulas whose size is always exponential in the number of eliminated atoms. In contrast, the method described in \Cref{sec:cip-by-dnf} does not increase the size of the involved formula. However, it requires these formulas to be in DNF, and transforming arbitrary formulas into DNF can lead to an exponential increase in the size of the formula. Other methods for interpolation try to compute interpolants in a more goal-oriented manner, with the aim of achieving efficiency in many practical cases,
even though so far no method is known that can compute interpolants whose size is guaranteed to be sub-exponential, and there might not even exist one (this is discussed in more detail in \Cref{sec:size-of-interpolants}).

One such method %
uses resolution. Resolution is a well-known method for deciding satisfiability of propositional formulas in CNF\red{~\cite{resolution:handbook:2001,kleinebuening:lettmann:1999}}. As mentioned in \Cref{sec:preliminaries}, it is possible to transform any propositional formula into a conservative extension in CNF in linear time. \red{In particular, many structure-preserving transformations such as the well-known Tseitin encoding} produce formulas that are not only equi-satisfiable, but also conservative extensions\red{~\cite{Baaz2001,Tseitin1968}}, which thanks to the following result is sufficient for computing interpolants.
\red{Since we will focus on Craig separators rather than Craig interpolants in this section, we directly formulate it for uniform interpolants and Craig separators.}
\begin{lemma}
 Let $\varphi$, $\psi$ be formulas such that $\varphi\wedge\psi\models\bot$, let $\sigma\subseteq \text{sig}(\varphi)$, and let $\varphi'$ and $\psi'$ be conservative extensions of $\varphi$ and $\psi$ that share no atoms that are not also shared by $\varphi$ and $\psi$. Then,
 every uniform %
 $\sigma$-interpolant for $\varphi'$ is also a uniform $\sigma$-interpolant for $\varphi$, and every Craig separator for $\varphi',\psi'$ is also a Craig separator for $\varphi,\psi$.
\end{lemma}
\begin{proof}
 The first claim follows directly from the definitions. 
 Now let $\chi$ be a Craig separator of $\varphi',\psi'$, which means that $\sig(\chi)\subseteq\sig(\varphi')\cap\sig(\psi')$, $\varphi'\models\chi$ and $\chi\wedge\psi'\models\bot$. By our assumption on the signatures of $\varphi',\psi'$, we have  $\sig(\chi)\subseteq\sig(\varphi)\cap\sig(\psi)$. By the definition of conservative extension, this implies $\varphi\models\chi$ and $\psi\models\neg\chi$, which means that $\chi$ is also a Craig separator for $\varphi,\psi$.
\end{proof}

Resolution for propositional logic operates on formulas in CNF, which in this
context are seen as sets
of clauses. Its central inference rule is the resolution rule:

\[
 \dfrac{C_1\vee p\qquad C_2\vee\neg p}{C_1\vee C_2}
\]

Since we represent clauses as sets, we can write this rule also as follows, highlighting the fact that the order of literals is not relevant and duplicates are silently removed,
and where we assume that $p\in C_1$ and $\neg p\in C_2$:
\[
 \dfrac{C_1\qquad C_2}{(C_1\setminus\{p\})\cup (C_2\setminus\{\neg p\})}
\]

The conclusion of the resolution rule is called \emph{resolvent} of
  the \name{parent clauses} $C_1$ and $C_2$, and we say
that it was derived by \emph{applying resolution upon $p$ to $C_1$ and $C_2$},
  the atom that has been eliminated in the resolvent.

It is well-known that resolution is sound and \emph{refutationally complete}, which means that we can derive the \emph{empty clause} from a set $N$ of clauses by a sequence of resolution steps iff $N$ is unsatisfiable. This also holds for a restricted version called \emph{ordered resolution}: here, we assume a linear order $\prec$ on atoms, and apply resolution only upon literals whose atom is maximal wrt. $\prec$ within the respective clauses.

Resolution can also be used to compute Craig interpolants and uniform interpolants, and in both cases, one may argue that resolution operates in a more goal-oriented fashion than the approaches we have seen so far. We first discuss \emph{uniform interpolation}, as this method is easier.

\subsection{Uniform Interpolation Using Resolution}
\label{sec-resolution-uniform-interpolation}

\red{Our method for computing uniform interpolants using resolution is based on the SCAN approach introduced for
first-order logic in~\cite{scan:1992}.}
For formulas in CNF, uniform interpolants can be computed by step-wise application of the following theorem.
\begin{theorem}\label{the:ui-resolution}
 Let $\Phi$ be a formula in CNF, $p$ an atom, and $\Psi$
 be obtained from $\Phi$ by
 \begin{enumerate}
  \item applying resolution exhaustively upon $p$ to all clauses in $\Phi$,
  \item removing all clauses that contain $p$.
 \end{enumerate}
  Then, $\Psi\equiv\exists p.\Phi$.
\end{theorem}

Before we prove the theorem, we illustrate on an example how it is used to compute uniform interpolants.

\begin{example}\label{ex:ui-resolution}
 Consider the formula
 \[
  \varphi = \neg(d\wedge e)\wedge(a\rightarrow d)\wedge(a\vee c)\wedge e .
 \]
 We want to compute a uniform $\{a,d\}$-interpolant for $\varphi$. We first compute the CNF $\Phi$ equivalent to $\varphi$, resulting in the following clauses:

\pagebreak 
 \begin{align*}
  &1.\ \neg d\vee\neg e  && \textit{input clause} \\
  &2.\ \neg a\vee d      && \textit{input clause} \\
  &3.\ a\vee c           && \textit{input clause} \\
  &4.\ e                 && \textit{input clause} %
 \intertext{
 We first compute a propositional formula equivalent to $\exists d.\Phi$ by applying resolution upon $d$:
 }
  & 5.\ \neg a\vee \neg e && \textit{ resolvent of 1, 2 upon $d$}%
 \intertext{
 $\exists d.\Phi$ is equivalent to the conjunction of Clauses~3--5,
 the clauses that do not contain $d$. We continue by computing the propositional formula equivalent to
 $\exists a.\exists d.\Phi$, for which we now have to apply resolution upon $a$ to the remaining clauses:
 }
  &6.\ c\vee \neg e &&\textit{ resolvent of 3, 5 upon $a$}%
 \end{align*}
 After removing all clauses that contain $a$, we obtain the uniform $\{a,d\}$-interpolant for $\varphi$,
 which is the conjunction of Clause~4 and~6,
 namely $e\wedge(c\vee \neg e)\equiv e\wedge c$.
\lipicsEnd
\end{example}

\begin{proof}[Proof of \Cref{the:ui-resolution}]
 We can prove the theorem in two ways. While the first proof is more straightforward, the
 second one uses an idea that is also used for more expressive logics, but is much easier to
 follow in the case of propositional logic (e.g. see \refchapter{chapter:kr}).

\subparagraph{Variant 1: Using quantifier elimination.}
 Assume that no clause contains both $p$ and $\neg p$. (Since such clauses 
 are tautologies, we can safely remove them.)
 By factoring out occurrences of $p$ and $\neg p$ respectively, we can transform
 $\Phi$ into an equivalent formula of the following form,
 where $\Phi_1$, $\Phi_2$ and $\Phi_3$ are formulas in CNF that contain no
 occurrences of~$p$:
 \begin{align}
  \Phi_1\wedge (\Phi_2\vee p)\wedge (\Phi_3\vee\neg p)
 \end{align}
 If we use the quantifier elimination technique from \Cref{sec:cip-by-quantifier-elimination}
 on $p$ in this formula, we obtain
 the following formula that is equivalent to $\exists p.\Phi$
 \begin{align*}
    & \left(\Phi_1\wedge (\Phi_2\vee \top)\wedge (\Phi_3\vee\neg \top)\right)\vee\left(\Phi_1\wedge (\Phi_2\vee \bot)\wedge (\Phi_3\vee\neg \bot)\right)\\
    \equiv & \red{(\Phi_1\wedge\top\wedge\Phi_3)\vee(\Phi_1\wedge\Phi_2\wedge\top)}\\
    \equiv &\ \Phi_1\wedge (\Phi_3\vee \Phi_2)
 \end{align*}
 \red{We can transform $(\Phi_3\vee\Phi_2)$ back into CNF using distributivity: the resulting formula is a conjunction
 over all clauses that can be obtained by combining a clause from $\Phi_3$ with another clause from $\Phi_2$.
 Note that those clauses are exactly those that we obtain when applying resolution upon $p$ in the original
 formula, so that the resulting set of clauses is indeed the} CNF formula $\Psi$ that is computed as in the theorem,
 so that indeed
 $\Psi\equiv\exists p.\Phi$.
 \qed

\subparagraph{Variant 2: Using ordered resolution.}
 For this proof, we use refutational completeness of ordered resolution, where we use an
 ordering $\prec_p$ that makes $p$ maximal.
 We need to show that
 \red{$\Psi$ is a uniform $\Sigma$-interpolant of $\Phi$ for
 $\Sigma=\sig(\Phi)\setminus\{p\}$. The first two conditions in \Cref{def:uniform-interpolant}
 hold because 1)~the resolvent is always entailed by its parent clauses, and 2)~we explicitly removed all
 occurrences of $p$. It remains to show that}
 for
 any formula $\varphi$ that does not contain $p$, $\Psi\models\varphi$ \red{if}
 $\Phi\models\varphi$.
 Since $\Phi\models\varphi$ iff
 $\Phi\wedge\neg\varphi\models\bot$, we can do so by
 showing that for any formula $\Theta$ in CNF s.t. $p\not\in\sig(\Theta)$,
 $\Phi\cup \Theta\models\bot$
 \red{implies} $\Psi\cup \Theta\models\bot$.
 \red{We can decide $\Phi\cup\Theta\models\bot$ using} ordered resolution with
 the ordering $\prec_p$. %
 In particular, this means that \red{we first} compute all resolvents upon $p$
 (which only occurs in $\Phi$, but not in $\Theta$), and
 then ignore occurrences of $p$ while we continue. Effectively,
 we thus first compute $\Psi$, and then continue on $\Psi\cup\Theta$.
 \red{Thus, if the empty clause can be derived from $\Phi\cup\Theta$, it can also be derived from $\Psi\cup\Theta$}.
\end{proof}

\subsection{Craig Interpolation Using Resolution}
\label{sec-resolution-craig-interpolation}

It is more convenient to look at Craig separators rather than Craig interpolants.
\red{To compute a Craig separator using resolution, we follow the method by Huang~\cite{huang:1995}.}
Let $\Phi$ and $\Psi$ be sets of clauses s.t. $\Phi\cup\Psi\models\bot$.
The idea is \red{to construct a Craig separator for $\Phi,\Psi$ by following the clauses used in the
\emph{resolution proof for $\Phi\cup\Psi\models\bot$}, by which we simply refer to the sequence
of resolution steps that are used to derive the empty clause from $\Phi\cup\Psi$}.
\red{For this, we
annotate each clause $\varphi$ in the proof---including the input clauses---}with a formula $\chi[\varphi]$
that serves as an \emph{intermediate separator}.

\begin{definition}\label{def:relational-interpolant}
 Let %
 $\theta$ be a clause.
 A formula $\chi$ is an \emph{intermediate separator of $\Phi$ and $\Psi$
 relative to $\theta$} if
 \begin{enumerate}
  \item $\sig(\chi)\subseteq\sig(\Phi)\cap\sig(\Psi)$,
  \item $\Phi\models \theta\vee\chi$, and
  \item $\Psi\models \theta\vee\neg\chi$.
 \end{enumerate}
\end{definition}

\red{The following corollary follows directly from the definition:}

\begin{corollary}\label{cor:ci-resolution}
 $\chi$ is a Craig separator for $\Phi$ and $\Psi$ iff it is an intermediate separator 
 of~$\Phi$ and $\Psi$ relative to the empty clause.
\end{corollary}

The clauses in the resolution proof are annotated inductively, starting from the
input clauses, and then following each resolution step.
This means that the intermediate separator for each derived clause is based on
the intermediate separators for the clauses that we apply resolution to.
How we build the next intermediate separator depends on the atom upon which we resolve.
Intuitively, the atom has to go into the formula if it is
from the common signature of $\Phi$ and $\Psi$. Otherwise,
we have to combine the annotations of the \red{parent clauses} in a clever way.

We first annotate the clauses in $\Phi\cup\Psi$.
We annotate every $\theta\in\Phi$ with $\chi[\theta]=\bot$,
and every $\theta\in \Psi$ with $\chi[\theta]=\top$.
If a clause $\theta$ is a resolvent of the parent clauses~$\theta_1$ and~$\theta_2$
upon $p$, where $\theta_1$ and $\theta_2$ are already annotated, we annotate $\theta$ with $\chi[\theta]$,
which is defined based on the origin of $p$:
\begin{enumerate}
 \item if $p\in\sig(\Phi)\cap\sig(\Psi)$, then
 $\chi[\theta]=(p\vee\chi[\theta_1])\wedge(\neg p\vee\chi[\theta_2])$;
 \item if $p\in\sig(\Phi)\setminus\sig(\Psi)$, then $\chi[\theta]=\chi[\theta_1]\vee\chi[\theta_2]$;
 \item if $p\in\sig(\Psi)\setminus\sig(\Phi)$, then $\chi[\theta]=\chi[\theta_1]\wedge\chi[\theta_2]$.
\end{enumerate}

We illustrate the technique with an example.
\begin{example}
 We use again the formula \[\varphi=\neg(d\wedge e)\wedge(a\rightarrow d)\wedge(a\vee c)\wedge e\] from
 \Cref{ex:ui-resolution}. %
 It turns out that $\varphi\models\psi$, where
 \[
  \psi = (b\rightarrow c)\wedge (d\rightarrow f)
 \]
 To compute a Craig interpolant for $\varphi,\psi$, we compute a Craig separator for $\varphi,\neg\psi$.
 We first need to transform these formulas into CNF.
 We already produced the CNF $\Phi$ of $\varphi$. For $\neg \psi$, its CNF $\Psi$ is obtained as follows:
 \begin{align*}
         &\neg\big((b\rightarrow c)\wedge (d\rightarrow f)\big) \\
  \equiv\ & (b\wedge\neg c)\vee ( d\wedge\neg f) \\
  \equiv\ & (b\vee d)\wedge(b\vee\neg f)\wedge(\neg c\vee d)\wedge(\neg c\vee\neg f)
 \end{align*}
 We derive the empty clause from $\Phi\cup\Psi$, and annotate each clause with
 an intermediate separator, which for readability we simplify as we go along.
 \begingroup
 \addtolength{\jot}{-2pt}
 \begin{align*}
   & \text{clause $\theta$} & \text{origin} && \text{ annotation $\chi[\theta] $}\\[-8pt]
   \cline{1-6}%
   1.\ & \neg d\vee\neg e    & \Phi && \bot \\%\label{eq:cir:1}\\
   2.\ & \neg a\vee d        & \Phi && \bot \\%\label{eq:cir:2}\\
   3.\ & a\vee c             & \Phi && \bot \\%\label{eq:cir:3}\\
   4.\ & e                   & \Phi && \bot \\[\bigskipamount]%
   5.\ & b\vee d             & \Psi && \top \\%\label{eq:cir:5}\\
   6.\ & b\vee\neg f         & \Psi && \top \\%\label{eq:cir:6}\\
   7.\ & \neg c\vee d        & \Psi && \top \\%\label{eq:cir:7}\\
   8.\ & \neg c\vee\neg f    & \Psi && \top \\[\bigskipamount]%
   9.\ & a\vee d    & \text{ resolvent of 3, 7}%
        && (c\vee\bot)\wedge(\neg c\vee\top)\equiv c \\%\label{eq:cir:9}\\
   10.\ & d          & \text{ resolvent of 9, 2}%
        && (c\vee\bot)\equiv c \\%\label{eq:cir:10} \\
   11.\ & \neg e     & \text{ resolvent of 10, 1}%
        && (d\vee c)\wedge(\neg d\vee\bot)\equiv(c\wedge \neg d)\\% \label{eq:cir:11} \\
   12.\ & \bot       & \text{ resolvent of 4, 11}%
        && (c\wedge \neg d)\vee\bot\equiv (c\wedge \neg d)
 \end{align*}
 \endgroup 
 We obtain that the Craig separator for $\Phi,\Psi$, and hence the Craig interpolant for
 $\varphi,\psi$  is $c\wedge \neg d$, and indeed, one can confirm that
 \begin{align*}
  & \neg(d\wedge e)\wedge(a\rightarrow d)\wedge(a\vee c)\wedge e \\
  \models\ & c\wedge \neg d\\
  \models\ & (b\rightarrow c)\wedge (d\rightarrow f).\\[-30pt]
 \end{align*}
 \lipicsEnd
\end{example}

It remains to show that the method is also correct. By \Cref{cor:ci-resolution}, we only need
to show that the annotations really correspond to intermediate separators.
\begin{lemma}\label{lem:intermediate}
 For every \red{clause $\theta$ in the resolution proof}, $\chi[\theta]$ is an intermediate separator for $\Phi$ and $\Psi$ relative to $\theta$.
\end{lemma}
\begin{proof}
 Our construction ensures that all formulas are within the signature $\sig(\Phi)\cap\sig(\Psi)$,
 which means we only need to show the other two conditions in
 \Cref{def:relational-interpolant}.
 For clauses directly occuring in $\Phi$ and $\Psi$, our construction makes these
 conditions trivially true: In particular, if $\theta\in\Phi$, then $\Phi\models\theta\vee\bot$ \red{and $\Psi\models\theta\vee\neg\bot$}, and if $\theta\in\Psi$, then \red{$\Phi\models\theta\vee\top$ and} $\Psi\models\theta\vee\neg\top$.

 For the derived clauses, we show Conditions~2 and~3 by
 induction.
 Let $\theta=\theta_1'\vee\theta_2'$ be the resolvent of two clauses 
 $\theta_1=\theta_1'\vee p$ and $\theta_2=\theta_2'\vee\neg p$, which are annotated by 
 formulas $\chi[\theta_1]$ and $\chi[\theta_2]$ for which we have already established 
 that they are intermediate interpolants. We distinguish the different cases in the construction of $\chi[\theta]$, and show that in each case, $\Phi\models\theta\vee\chi[\theta]$ and $\Psi\models\theta\vee\neg\chi[\theta]$.
 \begin{enumerate}
  \item $p\in\sig(\Phi)\cap\sig(\Psi)$.
  Then, $\chi[\theta]=(p\vee\chi[\theta_1])\wedge(\neg p\vee\chi[\theta_2])$.
  We first show that $\Phi\models\theta\vee\chi[\theta]$. Let $v$ be a model s.t. 
  $v\models\Phi$ and $v\not\models\theta$. By our inductive hypothesis,
  $v\models\theta_1'\vee p\vee\chi[\theta_1]$ and $v\models\theta_2'\vee\neg p\vee\chi[\theta_2]$. 
  Because $v\not\models\theta$ and $\theta=\theta_1'\vee\theta_2'$, neither $\theta_1'$ nor $\theta_2'$ can be satisfied in $v$, so that we obtain $v\models p\vee\chi[\theta_1]$ and 
  $v\models\neg p\vee\chi[\theta_2]$. This directly gives us $v\models\chi[\theta]$, which establishes that for all models $v$ s.t. $v\models\Phi$, $v\models\theta\vee\chi[\theta]$, and thus $\Phi\models\theta\vee\chi[\theta]$.
  
  We next show $\Psi\models\theta\vee\neg\chi[\theta]$, for which we again take a model $v$ 
  s.t. $v\models\Psi$ and $v\not\models\theta$. Our inductive hypothesis now gives us 
  $v\models\theta_1'\vee p\vee\neg\chi[\theta_1]$ and $v\models\theta_2'\vee\neg p\vee\neg\chi[\theta_2]$, and we can again use $v\not\models\theta$ to obtain 
  \begin{align}
  	v&\models p\vee\neg\chi[\theta_1] \label{eq:1}\\
  	v&\models\neg p\vee\neg\chi[\theta_2] \label{eq:2}
  \end{align} 
We need to show 
  $v\not\models(p\vee\chi[\theta_1])\wedge(\neg p\vee\chi[\theta_2])$. Assume that $v$ 
  does \red{satisfy} the first conjunct, that is,
  \begin{align}
  	v\models p\vee\chi[\theta_1] \label{eq:3}
  \end{align} 
From this, we can use \Cref{eq:1} to obtain $v\models p$, which together with \Cref{eq:2} gives us $v\models\neg\chi[\theta_2]$. 
  This implies that $v\not\models\neg p\vee\chi[\theta_2]$. Since this was a consequence of our assumption in \Cref{eq:3}, we obtain
  $v\not\models(p\vee\chi[\theta_1])\wedge(\neg p\vee\chi[\theta_2])=\chi[\theta]$, and thus $\Psi\models\theta\vee\neg\chi[\theta]$.

  \item $p\in\sig(\Phi)\setminus\sig(\Psi)$.
  By the inductive hypothesis, $\Phi\models\theta_1\vee\chi[\theta_1]$ and $\Phi\models\theta_2\vee\chi[\theta_2]$.
  We can weaken these disjunctions to $\theta_1\vee(\chi[\theta_1]\vee\chi[\theta_2])$ and
  $\theta_2\vee(\chi[\theta_1]\vee\chi[\theta_2])$,
  so that we can factor out $\chi[\theta_1]\vee\chi[\theta_2]=\chi[\theta]$,
  resulting in $\Phi\models(\theta_1\wedge \theta_2)\vee\chi[\theta]$.
  Because~$\theta$ was inferred from $\theta_1$ and $\theta_2$ through resolution, we have
  $\theta_1\wedge \theta_2\models \theta$, so that we obtain
  $\Phi\models\theta\vee\chi[\theta]$.

  Next, we show $\Psi\models\theta\vee\neg\chi[\theta]$. By inductive hypothesis,
  $\Psi\models\theta_1'\vee p\vee\neg\chi[\theta_1]$ and
  $\Psi\models\theta_2'\vee \neg p\vee\neg\chi[\theta_1]$.
  Since $p$ does not occur in $\Psi$, this implies
  $\Psi\models\theta_1'\vee\neg\chi[\theta_1]$ and
  $\Psi\models\theta_2'\vee\neg\chi[\theta_2]$. We can weaken $\theta_1'$ and $\theta_2'$ to $\theta_1'\vee\theta_2'=\theta$, to obtain $\Psi\models\theta\vee\neg\chi[\theta_1]$ and
  $\Psi\models\theta\vee\neg\chi[\theta_2]$. Factoring out $\theta$, we obtain
  $\Psi\models\theta\vee(\neg\chi[\theta_1]\wedge\neg\chi[\theta_2])=$
  $\theta\vee\neg(\chi[\theta_1]\vee\chi[\theta_2])=\theta\vee\neg\chi[\theta]$.

  \item $p\in\sig(\Psi)\setminus\sig(\Phi)$. This case is shown as the previous one, with the
  roles of $\Phi$ and $\Psi$ reversed.
  \qedhere
 \end{enumerate}
\end{proof}

\begin{lemma}\label{lem:resolution}
  Let $\Phi$, $\Psi$ be some set of clauses s.t. $\Phi\cup\Psi\models\bot$, and
  $\Theta$ be a sequence of clauses deriving the empty clause from $\Phi\cup\Psi$, where each
  clause is annotated as above. Then, the empty clause is annotated with a formula $\chi[\bot]$
  which is a Craig separator for $\Phi$ and $\Psi$.
\end{lemma}

A nice property of the construction is that we can construct from a given resolution proof for $\Phi\cup\Psi\models\bot$ a Craig separator whose size is polynomial in the number of inference steps in the proof. This is a desirable property of proof systems in general, that is commonly referred to as \emph{feasible interpolation}. A more detailed discussion on feasible interpolation is provided in \refchapter{chapter:proofcomplexity}. The interested reader might also consult~\cite{DBLP:journals/jsyml/Krajicek94,DBLP:journals/jsyml/Krajicek97,DBLP:journals/jsyml/Pudlak97}.
  
\begin{theorem}\label{the:resolution}
 Given a resolution proof for $\Phi\cup\Psi\models\bot$, we can in polynomial time construct a Craig separator for $\Phi,\Psi$ whose size is linearly bounded by the size of the proof.
\end{theorem}

\begin{proof}
By induction on the number of steps in the proof, we obtain that each clause
is annotated by a formula whose size is polynomial in the number of proof
steps that lead to this clause. For the input clauses, this holds trivially.
By analyzing the different cases for how we annotate new resolvents, we see
that each intermediate separator introduces at most five new subformulas.
Consequently, the (dag-)size of the computed separator is at most five times
the number of inference steps, and thus linearly bounded in the size of the
proof.
\end{proof}

Because of the case where we resolve upon some $p\in\sig(\Phi)\cap\sig(\Psi)$,
which introduces both $p$ and $\neg p$
into the Craig interpolant, the interpolants computed with this method
do not satisfy the Lyndon property. There are however alternative approaches, following the same idea,
but using a different way of annotating clauses, that do preserve the Lyndon property.
One such example is the interpolation system by McMillan~\cite{mcmillan:ipol:2003,mcmillan:ipol:2005}. Here, each input clause $\theta\in\Phi$
is annotated with the clause $\chi[\theta]$ that is obtained from $\theta$ by removing literals
that use atoms from $\sig(\Phi)\setminus\sig(\Psi)$, that is, keeping only those literals that are in the target signature.
Clauses in $\Psi$ are still annotated with $\top$, and clauses $\theta$ obtained  through resolution upon $p$ on clauses $\theta_1$, $\theta_2$ are annotated as before, except for the case where $p\in\sig(\Phi)\cap\sig(\Psi)$, in which
case we set $\chi[\theta]=\chi[\theta_1]\wedge\chi[\theta_2]$ (as in the case for $p\in\sig(\Psi)\setminus\sig(\Phi)$).

\subsection{Bibliographic and Historic Remarks}
\label{sec-historic-resolution}

\name{Resolution} was introduced in 1965 by J. Alan Robinson
\cite{robinson:1965} as a machine-suited proof method for first-order logic.
For propositional logic, it was, in the dual form for DNF, already established
since 1955 as the \name{consensus method} by Willard Van Orman Quine
\cite{quine:consensus:1955}, which is applied to convert a given DNF to a form
that consists of all its prime implicants (minimal conjunctive clauses without
complementary literals that imply the overall formula). The first appearance
of propositional resolvent/consensus construction is attributed to Archie
Blake in 1937 \cite{blake:thesis:1937,onblake:1986}.
Our resolution-based method for uniform interpolation from
Section~\ref{sec-resolution-uniform-interpolation} has ancestors in a more
general method for second-order quantifier elimination on the basis of
first-order logic developed by Wilhelm Ackermann in 1935
\cite{ackermann:1935}, which was rediscovered in 1992 by Dov M. Gabbay and
Hans Jürgen Ohlbach \cite{scan:1992}.
In its dual variation, our method for propositional logic is attributed to
Edward W. Samson and Rolf K. Mueller in 1955
\cite{samson:mueller:consensunselim:1955} (see
\cite[Sect.~7.2.2.2]{knuth:taocp:vol4:fas6}). They address the tautology
problem by eliminating all atoms from a given propositional DNF one by one
with consensus.
The system for Craig interpolation shown in
Section~\ref{sec-resolution-craig-interpolation} is called \name{HKPYM} by
Maria Paola Bonacina and Moa Johansson in their survey on ground interpolation
systems \cite{bonacina:2015:ground}, reflecting the initials of several
authors who apparently discovered it independently, Guoxiang Huang in 1995
\cite{huang:1995} (actually for first-order clauses), Jan Kraj{\'{\i}}cek
\cite{DBLP:journals/jsyml/Krajicek97} in 1997, and Pavel Pudl{\'{a}}k
\cite{DBLP:journals/jsyml/Pudlak97} also in 1997, as well as Greta Yorsh and
Madanlal Musuvathi \cite{yorsh:musvathi:2005} for contributing in 2005 a
completeness proof. In their survey, Bonacina and Johansson also present a
second resolution-based propositional interpolation system by Kenneth L.
McMillan from 2003 \cite{mcmillan:ipol:2003,mcmillan:ipol:2005} in a form that
allows easy comparison.

A comprehensive presentation of resolution in propositional logic is included
in the monograph \name{Propositional Logic -- Deduction and Algorithms} by
Hans Kleine Büning and Theodor Lettmann \cite{kleinebuening:lettmann:1999}. As
introductory presentations of resolution in general we recommend the classical
textbook by Chin-Liang Chang and Richard Char-Tung Lee
\cite{chang:lee:book:1973}, and the handbook article by Leo Bachmair and
Harald Ganzinger \cite{resolution:handbook:2001}.

\addtocontents{toc}{\pagebreak}

\newcommand{\tabrulename}[1]{\raisebox{-2.0ex}[0pt][0pt]{\small #1}}
\newcommand{\tableauform}{\mathcal{F}\!\mathit{orm}}

\newcommand{\nhphantom}[1]{\sbox0{#1}\hspace*{-\the\wd0}}
\newcommand{\nannot}[1]%
           {\text{\hspace{0.2em}{[}$#1${]}\nhphantom{\hspace{0.2em}{[}$#1${]}}}}

\newcommand{\closedmark}{\times}

\section{Craig Interpolants via Analytic Tableaux}
\label{sec-tableaux}

In this section we provide a further proof of the CIP, or actually the
stronger LIP, via a \emph{tableau} proof system. As for resolution, we adopt
the \emph{separator} view, since tableau systems are most commonly modeled as
\emph{refutational} systems that construct a tree-structured proof, a
\emph{closed tableau}, which certifies unsatisfiability of a given formula.
Again, as for resolution, computation of a Craig separator $\chi$ for $\varphi,
\psi$ proceeds in two phases: first, a proof of the unsatisfiability of
$\varphi \land \psi$ is obtained, which is here the tableau; second, a
separator $\chi$ is extracted from the proof. As we will see, extraction from
a tableau is a straightforward induction over its tree structure.
We describe our proof system in Section~\ref{sec-tab-proofsystem} and the
separator extraction in Section~\ref{sec-tab-separator}. The sources of the
underlying technique are the monographs by Fitting
\cite{fitting:book:foar:2nd} and Smullyan \cite{smullyan:book:1968}. Its
variation for sequent systems is known as \name{Maehara's method}
(see \refchapter{chapter:prooftheory}).

\subsection{A Tableau Proof System}
\label{sec-tab-proofsystem}

We consider a very simple analytic tableau system that operates on
propositional formulas in NNF.
It constructs a \defname{tableau}, a tree whose nodes are labeled by formulas.
For our system, this is an ordered binary tree. As common for tableau systems,
it is a refutation system, that is, it aims at showing unsatisfiability of a
given formula $\varphi$.
A tableau is said to be \defname{for} this given formula $\varphi$. The
tableau construction starts with a single-node tableau labeled with~$\varphi$.
A tableau branch is \defname{closed} if it contains a node labeled with
$\false$ or contains two nodes labeled with complementary literals, otherwise
it is called \defname{open}. A tableau is called \defname{closed} if all of
its branches are closed. The tableau construction repeatedly extends an open
branch of the tableau by applying one of the following two \defname{tableau
  expansion rules}.
\[
\begin{array}[t]{cl}
  \varphi_1 \land \varphi_2 & \; \tabrulename{$\alpha$} \\\cmidrule(){1-1}
  \varphi_1\\
  \varphi_2
\end{array}
\hspace{2cm}
\begin{array}[t]{cl}
  \varphi_1 \lor \varphi_2 & \; \tabrulename{$\beta$} \\\cmidrule(){1-1}  
  \varphi_1\; |\; \varphi_2
\end{array}
\]
Applying rule~$\alpha$ means selecting an open branch and a conjunction
$\varphi_1 \land \varphi_2$ on the branch, and then adding to the end of the
branch a node labeled with $\varphi_1$ and another node with $\varphi_2$ as
child of the latter one. Applying rule $\beta$ means selecting an open branch
and a disjunction $\varphi_1 \lor \varphi_2$ on the branch, and then adding to
the end of the branch a left child labeled with $\varphi_1$ and a right child
labeled with $\varphi_2$. The objective of the system is to construct a
tableau for~$\varphi$ that is closed, which indicates that $\varphi$ is
unsatisfiable.

\begin{example}We
  apply our tableau system to construct a closed tableau for the formula
  $\varphi = ((p \land \lnot q) \lor p) \land \lnot p$, the negation of
  $((p\imp q)\imp p)\imp p$ as NNF.

   \hspace{-18pt}
   \begin{tikzpicture}[scale=0.89,
      baseline=(a.north),
      sibling distance=4em,level distance=4em,
      parent anchor=south,
      child anchor=north,
      every node/.style = {transform shape}]
      \node (a) [anchor=north] {$\begin{array}{c}
          ((p \land \lnot q) \lor p) \land \lnot p\\
        \end{array}$}
      ;
  \end{tikzpicture}
  \raisebox{-1.5cm}{$\stackrel{\alpha}{\rightarrow}$}
  \begin{tikzpicture}[scale=0.89,
      baseline=(a.north),
      sibling distance=4em,level distance=4em,
      parent anchor=south,
      child anchor=north,
      every node/.style = {transform shape}]
      \node (a) [anchor=north] {$\begin{array}{c}
          ((p \land \lnot q) \lor p) \land \lnot p\\
          (p \land \lnot q) \lor p\\
          \lnot p\\
        \end{array}$}
      ;
  \end{tikzpicture}
  \raisebox{-1.5cm}{$\stackrel{\beta}{\rightarrow}$}  
  \begin{tikzpicture}[scale=0.89,
      baseline=(a.north),
      sibling distance=4em,level distance=4em,
      parent anchor=south,
      child anchor=north,
      every node/.style = {transform shape}]
      \node (a) [anchor=north] {$\begin{array}{c}
          ((p \land \lnot q) \lor p) \land \lnot p\\
          (p \land \lnot q) \lor p\\
          \lnot p\\
        \end{array}$}
      child { node [anchor=north] {$\begin{array}{c}
            p \land \lnot q\\
          \end{array}$}}
      child { node [anchor=north] {$\begin{array}{c}  
            p\\
          \end{array}$
        }
      };
  \end{tikzpicture}
  \raisebox{-1.5cm}{$\stackrel{\alpha}{\rightarrow}$}    
  \begin{tikzpicture}[scale=0.89,
      baseline=(a.north),
      sibling distance=4em,level distance=4em,
      parent anchor=south,
      child anchor=north,
      every node/.style = {transform shape}]
      \node (a) [anchor=north] {$\begin{array}{c}
          ((p \land \lnot q) \lor p) \land \lnot p\\
          (p \land \lnot q) \lor p\\
          \lnot p\\
        \end{array}$}
      child { node [anchor=north] {$\begin{array}{c}
            p \land \lnot q\\
            p\\
            \lnot q\\
          \end{array}$}}
      child { node [anchor=north] {$\begin{array}{c}  
            p\\
          \end{array}$}
      };
  \end{tikzpicture}
\lipicsEnd  
\end{example}

Our tableau system is \defname{sound}, that is, if it constructs a closed
tableau for $\varphi$, then $\varphi$ is indeed unsatisfiable.
One way to see this is via considering tableau construction as an
equivalence-preserving formula transformation, similar to DNF transformation:
We associate with a tableau a specific formula, the disjunction of, for each
branch, the conjunction of the branch's formula labels. For the initial
single-node tableau, the associated formula is then just the given formula
$\varphi$. The associated formula is preserved up to equivalence by
applications of the rules $\alpha$ and $\beta$. If the tableau is closed, then
each branch is closed, that is, contains a pair of complementary literals or
$\false$. Hence, for each branch the conjunction of its formula labels is then
equivalent to $\false$. Thus, also the overall formula associated with the
tableau, the disjunction of the formulas for the branches, is equivalent to
$\false$. It follows that if a tableau for $\varphi$ can be constructed that
is closed, then $\varphi \equiv \false$, i.e., $\varphi$ is indeed
unsatisfiable.

The system is also \defname{complete}, that is, if $\varphi$ is unsatisfiable,
then it constructs a closed tableau for $\varphi$. For proofs we refer to the
monographs by Smullyan \cite{smullyan:book:1968} or Fitting
\cite{fitting:book:foar:2nd}, where completeness of similar systems is
discussed.

\subsection{Craig-Lyndon Separator Extraction from a Closed Tableau}
\label{sec-tab-separator}

A closed tableau for a given conjunction $\varphi \land \psi$ allows to
extract a \emph{Craig-Lyndon separator}~$\chi$ for the conjuncts $\varphi,
\psi$.
To enable this, we enhance our tableau system by a means for keeping track of
provenance information, whether a formula in the tableau stems from $\varphi$
or from $\psi$. This is represented by wrapping formulas stemming from
$\varphi$ in $L(\_)$ (suggesting \name{left side}) and formulas stemming from
$\psi$ in $R(\_)$ (suggesting \name{right side}). We call formulas decorated
with such provenance information \defname{biased formulas}. The tableau
expansion rules $\alpha, \beta$  now come in \emph{biased} versions,
where provenance information is propagated.
\[
\begin{array}[t]{cl}
  L(\varphi_1 \land \varphi_2) & \; \tabrulename{$\alpha_L$} \\\cmidrule(){1-1}
  L(\varphi_1)\\
  L(\varphi_2)
\end{array}
\hspace{0.6cm}
\begin{array}[t]{cl}
  R(\varphi_1 \land \varphi_2) & \; \tabrulename{$\alpha_R$} \\\cmidrule(){1-1}
  R(\varphi_1)\\
  R(\varphi_2)
\end{array}
\hspace{0.6cm}
\begin{array}[t]{cl}
  L(\varphi_1 \lor \varphi_2) & \; \tabrulename{$\beta_L$} \\\cmidrule(){1-1}  
  L(\varphi_1)\; |\; L(\varphi_2)
\end{array}
\hspace{0.6cm}
\begin{array}[t]{cl}
  R(\varphi_1 \lor \varphi_2) & \; \tabrulename{$\beta_R$} \\\cmidrule(){1-1}  
  R(\varphi_1)\; |\; R(\varphi_2)
\end{array}
\]
The provenance indicators $L$ and $R$ are ignored in determining whether a
branch is closed. Thus, a branch can be closed in the following ways: by a pair
$L(\ell), L(\complit{\ell})$, by a pair $R(\ell), R(\complit{\ell})$, by a
pair $L(\ell), R(\complit{\ell})$, by $L(\false)$, and by $R(\false)$.

Let $S$ be a set of biased formulas. We say that a formula~$\chi$
\defname{separates} $S$, symbolically expressed as $S \tabsep \chi$, iff
$\chi$ is a Craig-Lyndon separator for $\bigwedge_{L(\theta) \in S} \theta,\;
\bigwedge_{R(\theta) \in S} \theta$. The following lemma gathers
the properties of $\tabsep$ that underly separator extraction.

\newcommand{\lemnum}[1]{\textbf{\textsf{\upshape{#1}}}}

\begin{lemma}
  \label{lem-tabsep}
  Let $S$ be a set of biased formulas. It then holds for literals $\ell$ and
  formulas $\varphi_1, \varphi_2, \chi, \chi_1, \chi_2$ that

  \sublemma{lem-tabsep-base-1} $S \cup \{L(\ell)\} \cup \{L(\complit{\ell})\} \tabsep \false$\par
  \sublemma{lem-tabsep-base-2}  $S \cup \{R(\ell)\} \cup \{R(\complit{\ell})\} \tabsep \true$\par
  \sublemma{lem-tabsep-base-3} $S \cup \{L(\ell)\} \cup \{R(\complit{\ell})\} \tabsep \ell$\par
  \sublemma{lem-tabsep-base-4} $S \cup \{L(\false)\} \tabsep \false$\par
  \sublemma{lem-tabsep-base-5} $S \cup \{R(\false)\}\tabsep \true$\par
  \sublemma{lem-tabsep-a-1} If $S \cup \{L(\varphi_1), L(\varphi_2)\} \tabsep \chi$, then
  $S \cup \{L(\varphi_1 \land \varphi_2)\} \tabsep \chi$\par
  \sublemma{lem-tabsep-a-2} If $S \cup \{R(\varphi_1), R(\varphi_2)\} \tabsep \chi$, then
  $S \cup \{R(\varphi_1 \land \varphi_2)\} \tabsep \chi$\par
  \sublemma{lem-tabsep-b-1} If
    $S \cup \{L(\varphi_1)\} \tabsep \chi_1$ and
    $S \cup \{L(\varphi_2)\} \tabsep \chi_2$, then
    $S \cup \{L(\varphi_1 \lor \varphi_2)\} \tabsep (\chi_1 \lor \chi_2)$\par
  \sublemma{lem-tabsep-b-2} If  
      $S \cup \{R(\varphi_1)\} \tabsep \chi_1$ and
      $S \cup \{R(\varphi_2)\} \tabsep \chi_2$, then
      $S \cup \{R(\varphi_1 \lor \varphi_2)\} \tabsep (\chi_1 \land \chi_2)$
\end{lemma}
Lemma~\ref{lem-tabsep} is easy to verify. For example,
\red{Lemma~}\ref{lem-tabsep-base-3} expresses that, for arbitrary sets $S$ of
biased formulas, the literal $\ell$ is a Craig-Lyndon separator for
$\bigwedge_{L(\theta) \in S} \theta \land \ell,\; \bigwedge_{R(\theta) \in S}
\theta \land \complit{\ell}$. In case $\ell$ is a positive literal $p$, this
holds, since $\bigwedge_{L(\theta) \in S} \theta \land p \entails p$,
$\bigwedge_{R(\theta) \in S} \theta \land \lnot p \land p \entails \false$,
and $\sigposneg(p) = \{\langle p, {+}\rangle\} \subseteq
\sigposneg(\bigwedge_{L(\theta) \in S} \theta \land p) \cap \sigposneg(\lnot
(\bigwedge_{R(\theta) \in S} \theta \land \lnot p))$. The case for $\ell$ a
negative literal $\lnot p$ holds analogously.

As basis for Craig-Lyndon separator extraction, we build a closed biased
tableau \defname{for $\varphi, \psi$} with the biased rules, starting from the
initial biased tableau
\[
\begin{array}{c}
  L(\varphi)\\
  R(\psi)
\end{array}
\]

Lemma~\ref{lem-tabsep} then straightforwardly suggests a recursive separator
extraction procedure. We start with the closed tableau and determine as base
cases for each of its branches a separator, according to one of the
Lemmas~\ref{lem-tabsep-base-1}--\ref{lem-tabsep-base-5}. Since each branch is
closed, it must match with at least one of these lemmas.

We then proceed by considering each application of an expansion rule in the
tableau construction in the \emph{reverse order} of their application. If the
last applied rule was $\alpha_L$, the branch after rule application matches
the if-condition of Lemma~\ref{lem-tabsep-a-1} and the shorter branch before
rule application matches the conclusion of the lemma. Hence, the formula
$\chi$, which separates the branch after rule application, also separates the
shorter branch before rule application. Analogously, $\alpha_R$ corresponds to
Lemma~\ref{lem-tabsep-a-2}.
If the last applied rule was $\beta_L$, we have two branches, matching the two
if-conditions of Lemma~\ref{lem-tabsep-b-1}, and the branch before rule
application matches the conclusion of the lemma. Hence, from the two formulas
$\chi_1$ and $\chi_2$, each separating one of the two branches after rule
application, we build the disjunction $\chi_1 \lor \chi_2$, which separates
the shorter common branch before rule application. Analogously, $\beta_R$
corresponds to Lemma~\ref{lem-tabsep-b-2}, where the separator for the shorter
branch before rule application is the conjunction $\chi_1 \land \chi_2$.

Finally, after having processed the very first rule application, we obtain a
formula~$\chi$ that separates $\{L(\varphi), R(\psi)\}$, the initial branch
with which we started the tableau construction. In other words, $\chi$ is the
desired Craig-Lyndon separator for $\varphi, \psi$, and the extraction
is completed.

If we have a given closed tableau for $\varphi \land \psi$ that is
``unbiased'', obtained with the plain proof system of
Section~\ref{sec-tab-proofsystem}, it can easily be preprocessed to a
\emph{biased} tableau for $\varphi,\psi$: The initial application of $\alpha$
that leads to $\varphi$ followed by $\psi$ is dropped, such that the proof
only starts with $\varphi$ followed by $\psi$. If this occurrence of $\varphi$
is replaced by $L(\varphi)$ and the occurrence of $\psi$ by $R(\varphi)$, then
all rule applications can be replaced by biased versions that propagate the
provenance information downwards.

\begin{example}
  \label{examp-tab-separator-steps} To illustrate
  Craig-Lyndon separator extraction, we start from a closed biased tableau for
  $\varphi,\psi$ with $\varphi = (p \land q) \land r$ and $\psi = (\lnot p \lor
  \lnot q) \land \lnot s$, where separator values for closed branches
  according to Lemma~\ref{lem-tabsep-base-3} are annotated at the leaves. Then
  we undo the extension steps that led to the closed tableau, propagating the
  separator values upwards according to the respective cases of
  Lemma~\ref{lem-tabsep}. We finally arrive at the initial tableau with two
  nodes $L(\varphi)$ and $R(\psi)$ and obtain $\chi = p \land q$ as separator
  for $\varphi, \psi$.

  \vspace{-5pt}
  \noindent\hspace{-0.2cm}
  \begin{tikzpicture}[scale=0.78,
      baseline=(a.north),
      sibling distance=5em,level distance=7em,
      parent anchor=south,
      child anchor=north,
      every node/.style = {transform shape}]
    \node (a) [anchor=north] {$\begin{array}{c}
        L((p \land q) \land r)\\
        R((\lnot p \lor \lnot q) \land \lnot s)\\
        R(\lnot p \lor \lnot q)\\
        R(\lnot s)\\
        L(p \land q)\\
        L(r)\\
        L(p)\\
        L(q)\\
      \end{array}$}
    child { node [anchor=north] {$\begin{array}{c}
          R(\lnot p)\nannot{p}\\
        \end{array}$}}
    child { node [anchor=north] {$\begin{array}{c}
          R(\lnot q)\nannot{q}\\
        \end{array}$}};
  \end{tikzpicture}
  \raisebox{-1.4cm}{$\!\!\!\!\!\stackrel{\beta_R}{\leftarrow}\!\!\!\!\!$}
  \begin{tikzpicture}[scale=0.78,
      baseline=(a.north),
      sibling distance=5em,level distance=7em,
      parent anchor=south,
      child anchor=north,
      every node/.style = {transform shape}]
    \node (a) [anchor=north] {$\begin{array}{c}
        L((p \land q) \land r)\\
        R((\lnot p \lor \lnot q) \land \lnot s)\\
        R(\lnot p \lor \lnot q)\\
        R(\lnot s)\\
        L(p \land q)\\
        L(r)\\
        L(p)\\
        L(q)\nannot{p \land q}\\
      \end{array}$};
  \end{tikzpicture}
  \raisebox{-1.4cm}{$\!\!\!\!\!\stackrel{\alpha_L}{\leftarrow}\!\!\!\!\!$}
  \begin{tikzpicture}[scale=0.78,
      baseline=(a.north),
      sibling distance=5em,level distance=7em,
      parent anchor=south,
      child anchor=north,
      every node/.style = {transform shape}]
    \node (a) [anchor=north] {$\begin{array}{c}
        L((p \land q) \land r)\\
        R((\lnot p \lor \lnot q) \land \lnot s)\\
        R(\lnot p \lor \lnot q)\\
        R(\lnot s)\\
        L(p \land q)\\
        L(r)\nannot{p \land q}\\
      \end{array}$};
  \end{tikzpicture}
  \raisebox{-1.4cm}{$\!\!\!\!\stackrel{\alpha_L}{\leftarrow}\!\!\!\!$}
  \begin{tikzpicture}[scale=0.78,
      baseline=(a.north),
      sibling distance=5em,level distance=7em,
      parent anchor=south,
      child anchor=north,
      every node/.style = {transform shape}]
    \node (a) [anchor=north] {$\begin{array}{c}
        L((p \land q) \land r)\\
        R((\lnot p \lor \lnot q) \land \lnot s)\\
        R(\lnot p \lor \lnot q)\\
        R(\lnot s)\nannot{p \land q}\\
      \end{array}$};
  \end{tikzpicture}
  \raisebox{-1.4cm}{$\;\stackrel{\alpha_R}{\leftarrow}\!\!\!\!\!\!\!\!\!$}
  \begin{tikzpicture}[scale=0.78,
      baseline=(a.north),
      sibling distance=5em,level distance=7em,
      parent anchor=south,
      child anchor=north,
      every node/.style = {transform shape}]
    \node (a) [anchor=north] {$\begin{array}{c}
        L((p \land q) \land r)\\
        R((\lnot p \lor \lnot q) \land \lnot s)\nannot{p \land q}\\
      \end{array}$};
  \end{tikzpicture}
  \vspace{-10pt}\par\lipicsEnd
\end{example}

\begin{example}
  \label{examp-tab-separator-complex} 
  Figure~\ref{fig-tab-separator-complex} shows a more complex example of the
  separator extraction. Differently from
  Example~\ref{examp-tab-separator-steps}, we do not exhibit the undone
  tableau extension steps but simultaneously annotate all nodes that were
  leaves during the tableau construction with the respective separators. The
  given formulas are $\varphi = d \land (\lnot d \lor (a \land (b \lor c)))$
  and $\psi = (b \land (\lnot b \lor (\lnot a \land \lnot e))) \lor (\lnot b
  \land \lnot c)$. The obtained separator is $\false \lor ((\true \land a)
  \land (b \lor c))$, which simplifies to $a \land (b \lor c)$.
\lipicsEnd
\end{example}

\begin{example}
  \label{examp-tab-separator-different} Two different
  proofs of the unsatisfiability of $\varphi \land \psi$ can lead to different
  extracted separators. Here is a simple example. The given formulas are
  $\varphi = a \land (b \land c)$ and $\psi = (\lnot a \land \lnot d) \land
  (\lnot b \land \lnot d)$. As separators we obtain $a$ and $b$, depending on
  the closed biased tableau for $\varphi,\psi$. As in
  Example~\ref{examp-tab-separator-complex}, we simultaneously annotate all
  nodes that were leaves during the tableau construction with the respective
  separators.

  \medskip
  
$\begin{array}{c}
L(a \land  (b \land  c))\\
R((\lnot a \land  \lnot d) \land  (\lnot b \land  \lnot d))\nannot{a}\\
L(a)\\
L(b \land c)\nannot{a}\\
R(\lnot a \land  \lnot d)\\
R(\lnot b \land  \lnot d)\nannot{a}\\
R(\lnot a)\\        
R(\lnot d)\nannot{a}
\end{array}$
  \hspace{1.5cm}
$\begin{array}{c}
L(a \land  (b \land  c))\\
R((\lnot a \land  \lnot d) \land  (\lnot b \land  \lnot d))\nannot{b}\\
L(a)\\
L(b \land c)\nannot{b}\\
R(\lnot a \land  \lnot d)\\
R(\lnot b \land  \lnot d)\nannot{b}\\
R(\lnot b)\\        
R(\lnot d)\nannot{b}
  \end{array}$
\vspace{-10pt}\par\lipicsEnd
\end{example}

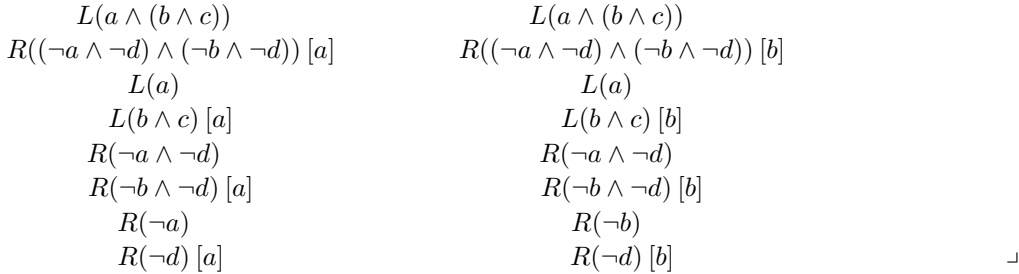
\begin{figure}[t]
  \centering

  \begin{tikzpicture}[%
      scale=0.85,
      baseline=(a.north),
      sibling distance=12em,level distance=6em,
      parent anchor=south,
      child anchor=north,
      every node/.style = {transform shape}]
      \node (a) [anchor=north] {$\begin{array}{c}
          L(d \land (\lnot d \lor (a \land (b \lor c))))\\
          R((b \land (\lnot b \lor (\lnot a \land \lnot e)))
          \lor (\lnot b \land \lnot c))%
          \nannot{\false \lor ((\true \land a) \land (b \lor c))}\\          
          L(d)\\
          L(\lnot d \lor (a \land (b \lor c)))%
          \nannot{\false \lor ((\true \land a) \land (b \lor c))}\\
        \end{array}$}
            [sibling distance=8em]      
      child { node [anchor=north] {$\begin{array}{c}
            L(\lnot d)\nannot{\false}\\
          \end{array}$}}
      child { node [anchor=north] {$\begin{array}{c}  
            L(a \land (b \lor c))\nannot{(\true \land a) \land (b \lor c)}\\
            L(a)\\
            L(b \lor c)\nannot{(\true \land a) \land (b \lor c)}\\
          \end{array}$}
        [sibling distance=16em]        
        child { node [anchor=north] {$\begin{array}{c}  
              R(b \land (\lnot b \lor (\lnot a \land \lnot e)))\nannot{\true \land a}\\
              R(b)\\
              R(\lnot b \lor (\lnot a \land \lnot e))\nannot{\true \land a}\\
            \end{array}$}
          [sibling distance=8em]
          child { node [anchor=north] {$\begin{array}{c}  
                R(\lnot b)\nannot{\true}\\
              \end{array}$}}
          child { node [anchor=north] {$\begin{array}{c}
                R(\lnot a \land \lnot e)\nannot{a}\\
                R(\lnot a)\\
                R(\lnot e)\nannot{a}\\
              \end{array}$}}}
        child { node [anchor=north] {$\begin{array}{c}
              R(\lnot b \land \lnot c)\nannot{b \lor c}\\
              R(\lnot b)\\
              R(\lnot c)\nannot{b \lor c}\\
            \end{array}$}
          [sibling distance=8em]
          child { node [anchor=north] {$\begin{array}{c}  
                L(b)\nannot{b}\\
              \end{array}$}}
          child { node [anchor=north] {$\begin{array}{c}
                L(c)\nannot{c}\\
              \end{array}$}}}
      };
  \end{tikzpicture}

  \caption{A more complex example of Craig-Lyndon separator extraction,
    discussed in Example~\ref{examp-tab-separator-complex}.}
    
\label{fig-tab-separator-complex}
\end{figure}
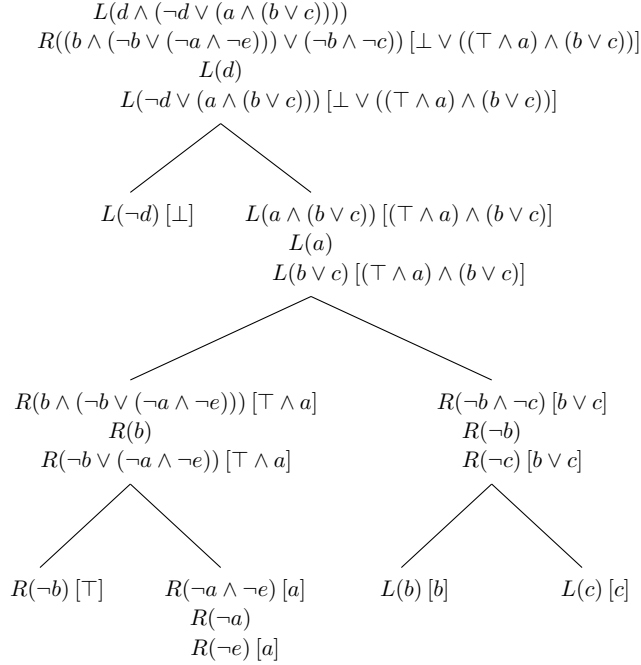

We can now summarize the results of this section in a proof statement of the
Craig-Lyndon Interpolation Property (LIP, Theorem~\ref{thm-cip-lyndon}),
expressed in terms of Craig-Lyndon separation.

\begin{proof}[Proof (Craig-Lyndon Interpolation Property (LIP))]
  Let $\varphi, \psi$ be propositional formulas such that $\varphi \land \psi$
  is unsatisfiable. We assume that both formulas are in NNF, which is
  w.l.o.g.\ as conversion to an equivalent NNF is always possible in linear
  time. Since the tableau proof system from Section~\ref{sec-tab-proofsystem}
  is complete, it yields a closed tableau for $\varphi \land \psi$. By adding
  provenance indicators it can be converted to a biased tableau for
  $\varphi,\psi$. Based on Lemma~\ref{lem-tabsep}, from the biased tableau we
  can extract a formula~$\chi$ such that $\{L(\varphi), R(\psi)\} \tabsep
  \chi$, that is, $\chi$ is a Craig-Lyndon separator for $\varphi, \psi$.
\end{proof}

This proof permits the following corollary about computational effort as well
as the shape and size of the separator in relation to the size of the
underlying proof, i.e., the tableau.

\begin{corollary}
  The time for extracting a Craig-Lyndon separator from a closed tableau is
  polynomial in the number of tableau nodes. The separator is in NNF and its
  size is at most linear in the number of tableau nodes.
\end{corollary}

That the separator extraction is in polynomial time can be seen as follows.
The extraction is performed in steps that each correspond to a dedicated
tableau node: to a leaf for each closed branch
(Lemmas~\ref{lem-tabsep-base-1}--\ref{lem-tabsep-base-5}), and to an inner
node for each application of an extension rule
(Lemmas~\ref{lem-tabsep-a-1}--\ref{lem-tabsep-b-2}). Within each of these
steps, a separating formula is determined according to the respective lemma,
which is clearly polynomial in the number of tableau nodes, as it involves
matching of only one or two branches with the preconditions of the respective
lemma, and the effort for constructing the separating formula is at most
building the disjunction or conjunction of two given formulas.

The NNF shape and the size bound follow since the extracted separator is built
by disjunction and conjunction from exactly one occurrence of a literal or
truth value constant for each tableau leaf.

\subsection{Bibliographic and Historic Remarks}
\label{sec-historic-tableaux}

Analytic tableaux became widely known with Raymond Smullyan's compact
monograph on first-order logic from 1968 \cite{smullyan:book:1968}. They are
related to earlier systems by, among others, Jaakko Hintikka 1955
\cite{hintikka:formandcontent:1955} and Evert W. Beth 1959
\cite{beth:book:foundations:1959}.
For a comprehensive historic account, see Melvin Fitting's introduction
\cite{handbook:tableaux:intro} to the \name{Handbook of Tableau Methods}
\cite{handbook:tableaux}. A monograph by Fitting \cite{fitting:book:foar:2nd}
and a handbook article by Reiner Hähnle \cite{handbook:ar:haehnle:2001} give
presentations of tableaux that take aspects of automated reasoning into
account.
Our interpolant extraction from analytic tableaux is based on the monograph by
Fitting from 1996 \cite{fitting:book:foar:2nd} and the earlier monograph by
Smullyan \cite{smullyan:book:1968}. Fitting specifies it by means of a rule
system.
Similar interpolation methods were presented on the basis of semantic tableaux
by Beth in 1959 \cite{beth:book:foundations:1959} and on the basis of sequent
calculi by Shoji Maehara in 1960 \cite{maehara:1960,takeuti:book:1987} and
also by Smullyan in 1968 \cite{smullyan:book:1968}.

\FloatBarrier

\section{The Size of Interpolants}
\label{sec:size-of-interpolants}
The algorithms computing Craig interpolants discussed above all output interpolants of exponential size in the size of the input formulas, in the worst case. The only exception is the DNF based algorithm which constructs interpolants by dropping subformulas from the left-hand side of the input entailment. However, since the transformation of an arbitrary formula into a logically equivalent formula in DNF can lead to an exponential increase in the size of the formula, also the DNF based algorithm outputs exponential size interpolants when applied to arbitrary input formulas. In fact, no algorithm computing Craig interpolants of polynomial size is known and it is widely conjectured that none exists. Proving this conjecture is hard, however, and would solve longstanding open problems in complexity theory. Even for uniform interpolants, no exponential (and not even superpolynomial) lower bounds are known. In this section, we discuss results that formalise this claim. \red{We also refer the reader to~\refchapter{chapter:proofcomplexity}, where many of the concepts used in this section are introduced in detail as part of a discussion of the role of Craig interpolation in proof complexity.}

Recall that we denote by $|\varphi|$ and $s(\varphi)$ the dag- and, respectively, tree-size of a propositional formula $\varphi$. It is conjectured that there is a superpolynomial gap between the dag-size and the tree-size of propositional formulas. This conjecture remains open~\cite{DBLP:journals/siamcomp/Rossman18}. The following observation states a close link between this conjecture and the tree-size of Craig interpolants. We say that propositional logic has \emph{polynomial tree-size Craig interpolants} if there is a polynomial function $f:\mathbb{N} \times \mathbb{N} \rightarrow \mathbb{N}$ such that for any $\varphi,\psi$ with $\varphi\models \psi$ there exists a Craig interpolant $\chi$ for $\varphi,\psi$ with $s(\chi) \leq f(|\varphi|,|\psi|)$.
\begin{theorem}
    If there is a superpolynomial gap between dag-size and tree-size, then
    propositional logic does not have polynomial tree-size Craig interpolants.
\end{theorem}
\begin{proof}
Assume there is a sequence of propositional formulas $(\varphi_{n})_{n\in \mathbb{N}}$ such that the tree-size of any equivalent propositional formulas 
grows superpolynomially compared to the dag-size of the formulas $\varphi_{n}$. Introduce for every subformula $\psi$ of $\varphi_{n}$ an atom $p_{\psi}$ and let $\psi_{n}$ be the conjunction of all 
\[
p_{q} \leftrightarrow q, \quad p_{\top} \leftrightarrow \top, \quad p_{\bot} \leftrightarrow \bot, \quad p_{\psi \wedge \chi}\leftrightarrow p_{\psi}\wedge p_{\chi},\quad
p_{\psi \vee \chi}\leftrightarrow p_{\psi}\vee p_{\chi}, \quad p_{\neg \psi} \leftrightarrow \neg p_{\psi}
\]
with $q$ ranging over atoms in $\varphi_{n}$, and $\psi\wedge \chi$, $\psi \vee\chi$, and $\neg \psi$ ranging over subformulas of $\varphi_{n}$. Let $\sigma_{n}= \sig(\varphi_{n})$. Then $\psi_{n} \models p_{\varphi_{n}}\leftrightarrow \varphi_{n}$ and so we have explicit $\sigma_{n}$-definitions of $p_{\varphi_{n}}$ under $\psi_{n}$. There cannot be polynomial tree-size explicit $\sigma_{n}$-definitions of $p_{\varphi_{n}}$ under $\psi_{n}$ since these would be propositional formulas equivalent to $\varphi_{n}$ of polynomial tree-size. It follows from the proof of Theorem~\ref{thm-beth} that Craig interpolants can also not be of polynomial tree-size.
\end{proof}
In what follows, we only consider the dag-size of a formula. So we say that propositional logic has \emph{polysize Craig interpolants} if there is a polynomial function $f:\mathbb{N} \times \mathbb{N} \rightarrow \mathbb{N}$ such that for any $\varphi,\psi$ with $\varphi\models \psi$ there exists a Craig interpolant $\chi$ for $\varphi,\psi$ with $|\chi| \leq f(|\varphi|,|\psi|)$. 
We say that propositional logic has \emph{polysize uniform interpolants} if there is a polynomial function $f:\mathbb{N} \rightarrow \mathbb{N}$ such that for any $\varphi$ and ordering ${\bf p}=p_{1},\ldots,p_{n}$ of any subset of $\sig(\varphi)$, there exists a propositional formula $\chi$ equivalent to $\exists {\bf p}.\varphi$ with $|\chi| \leq f(|\varphi|)$, where $\exists {\bf p}.\varphi$ stands for $\exists p_{1}\cdots \exists p_{n}.\varphi$. 

We start with uniform interpolants and show
the following equivalence.
\begin{theorem}~\label{thm:uniformequiv}
Propositional logic has polysize uniform interpolants iff ${\bf NP}\subseteq {\bf P}_{{\bf /poly}}$.
\end{theorem}
Before proving Theorem~\ref{thm:uniformequiv},
we explain the notation and concepts used in its formulation. We hope to convince a reader not familiar with computational complexity, in particular circuit complexity, that it is essentially a complexity theoretic reformulation of the definition of uniform interpolants. Note that the inclusion ${\bf NP}\subseteq {\bf P}_{{\bf /poly}}$ (stating that every problem in the complexity class ${\bf NP}$ is also in the complexity class ${\bf P}_{{\bf /poly}}$) is a long-standing open problem in complexity theory. We comment on the status of this problem below, once the notation is introduced. 

We introduce the relevant complexity classes,
in particular those defined using computations with Boolean circuits. We refer the reader to~\cite{DBLP:books/daglib/0023084} for further details. For our purposes, we can identify a Boolean circuit with a propositional formula. Let ${\bf p}=p_{1},\ldots,p_{n}$ be an ordering of the atoms in a propositional formula $\varphi$. Then $\varphi$ computes the function $C_{\varphi}:\{0,1\}^{n}\rightarrow \{0,1\}$ defined by setting $C_{\varphi}(v(p_{1}),\ldots,v(p_{n}))=v(\varphi)$, with $v$ any model mapping the atoms in ${\bf p}$ to $\{0,1\}$. Call $n$ the \emph{arity} of $\varphi$.

As usual for decision problems in computational complexity, we assume these are encoded as membership problems for subsets $S$ of $\{0,1\}^{\ast}$, the set of words of arbitrary length over the alphabet $\{0,1\}$. For the definition of the complexity classes {\bf P}, {\bf NP}, and {\bf coNP} via Turing machines we refer the reader to~\cite{DBLP:books/daglib/0023084}. To define the complexity of $S\subseteq \{0,1\}^{\ast}$ in terms of computations with Boolean circuits, we require families $(\varphi_{n})_{n\in \mathbb{N}}$ of propositional formulas, where $\varphi_{n}$ has arity $n$. Then $(\varphi_{n})_{n\in \mathbb{N}}$ \emph{decides $S$} if for all $n\in \mathbb{N}$ and $t_{1}\cdots t_{n}\in \{0,1\}^{n}$, $C_{\varphi_{n}}(t_{1},\ldots,t_{n})=1$ iff $t_{1}\cdots t_{n}\in S$. Note that the classes {\bf P}, {\bf NP}, and {\bf coNP} are defined using a single Turing machine that accepts inputs of arbitrary length. In contrast, the complexity classes defined using Boolean circuits require families of circuits, one for each arity $n\in \mathbb{N}$.

Now $S$ is in the complexity class ${\bf P}_{{\bf /poly}}$ if there exists a family $(\varphi_{n})_{n\in \mathbb{N}}$ of formulas deciding $S$ and a polynomial function $f:\mathbb{N} \rightarrow \mathbb{N}$ such that $|\varphi_{n}|\leq f(n)$ for all $n\in \mathbb{N}$. To define the class ${\bf NP}_{{\bf /poly}}$ we require quantified Boolean formulas of the form $\varphi=\exists q_{1}\cdots \exists q_{m}.\psi$ with $\psi$ a propositional formula. Any such quantified Boolean formula with additional non-quantified atoms $p_{1},\ldots,p_{n}$ computes the function $C_{\varphi}:\{0,1\}^{n} \rightarrow \{0,1\}$ defined by setting $C_{\varphi}(v(p_{1}),\ldots,v(p_{n}))=v(\varphi)$ with $v$ any model mapping the atoms $p_{1},\ldots,p_{n}$ to $\{0,1\}$. The \emph{arity} of $\varphi$ is $n$. Intuitively, to decide whether $t_{1}\cdots t_{n}\in \{0,1\}^{n}$ is in a set $S$ we guess a value $u_{1}\cdots u_{m}\in \{0,1\}^{m}$ (a certificate) and check whether $v\models \psi$ for the model $v$ with $v(p_{1})=t_{1},\ldots,v(p_{n})=t_{n},v(q_{1})=u_{1},v(q_{m})=u_{m}$. We can now generalise the definitions
for ${\bf P}_{{\bf /poly}}$ to ${\bf NP}_{{\bf /poly}}$ in the obvious way. A family $(\varphi_{n})_{n\in \mathbb{N}}$ with $\varphi_{n}=\exists q_{1}\cdots \exists q_{m_{n}}.\psi_{n}$ \emph{decides $S$} if for all $n\in \mathbb{N}$ and $t_{1}\cdots t_{n}\in \{0,1\}^{n}$, $C_{\varphi_{n}}(t_{1},\ldots,t_{n})=1$ iff $t_{1}\cdots t_{n}\in S$. $S$ is in the complexity class ${\bf NP}_{{\bf /poly}}$ if there exists such a family $(\varphi_{n})_{n\in \mathbb{N}}$ deciding $S$ and a polynomial function $f:\mathbb{N} \rightarrow \mathbb{N}$ such that $|\psi_{n}|\leq f(n)$ for all $n\in \mathbb{N}$. ${\bf coNP}_{{\bf /poly}}$ is the complement of ${\bf NP}_{{\bf /poly}}$. The following inclusions hold~\cite{DBLP:books/daglib/0023084}.
\begin{theorem}\label{thm:inc}
${\bf P} \subseteq {\bf P}_{{\bf /poly}}$, ${\bf NP} \subseteq {\bf NP}_{{\bf /poly}}$,
and ${\bf coNP} \subseteq {\bf coNP}_{{\bf /poly}}$.
\end{theorem}
We return to Theorem~\ref{thm:uniformequiv}. As mentioned above, the inclusion
${\bf NP} \subseteq {\bf P}_{{\bf /poly}}$ is a long-standing open problem. It is regarded as very unlikely that this inclusion holds. In fact, by the Karp-Lipton Theorem, if ${\bf NP} \subseteq {\bf P}_{{\bf /poly}}$, then the polynomial hierarchy collapses at the second level which is regarded as unlikely~\cite{karp1980some,karp1982turing,DBLP:books/daglib/0023084}.
Conversely, if we could prove ${\bf NP}\not\subseteq {\bf P}_{{\bf /poly}}$, then
${\bf NP} \not= {\bf P}$ would follow since ${\bf P}\subseteq {\bf P}_{{\bf /poly}}$.
So we would have solved one of the most famous open questions in computer science. 

We are now in a position to prove Theorem~\ref{thm:uniformequiv}. Assume first that propositional logic has polysize uniform interpolants. Let $S\subseteq \{0,1\}^{\ast}$ be in ${\bf NP}$. By Theorem~\ref{thm:inc}, $S$ is in ${\bf NP}_{{\bf /poly}}$. Then, by definition, we find a family $(\exists {\bf q}_{n}.\psi_{n})_{n\in \mathbb{N}}$ deciding $S$ with $|\psi_{n}|$ bounded by a polynomial function.
Since we assume that polysize uniform interpolants exist we find polysize $\chi_{n}$ logically equivalent to $\exists {\bf q}_{n}.\psi_{n}$.
But then
$(\chi_{n})_{n\in \mathbb{N}}$ decides $S$ and so $S$ is in ${\bf P}_{{\bf /poly}}$.

Conversely, we show that ${\bf NP}\subseteq {\bf P}_{{\bf /poly}}$ implies that uniform interpolants of polynomial size exist.
Let $\exists P_{n}$ denote the set of formulas of the form $\varphi=\exists q_{1}\cdots\exists q_{m}.\psi$ with $\psi$ a propositional formula and $\sig(\varphi)=\{p_{1},\ldots,p_{n}\}$ the remaining atoms in $\varphi$ distinct from $q_{1},\ldots,q_{m}$. It suffices to show that ${\bf NP}\subseteq {\bf P}_{{\bf /poly}}$ implies that
formulas in $\exists P=\bigcup_{n\in \mathbb{N}}\exists P_{n}$ are equivalent to polysize propositional formulas. Let $w_{\varphi}\in \{0,1\}^{\ast}$ be an encoding of formulas $\varphi\in \exists P$. We regard every $w\in \{0,1\}^{n}$ as a model with $w(p_{i})=w_{i}$ for $1\leq i \leq n$. Consider the set
\[
S_{n}= \{ w_{\varphi}w'\in \{0,1\}^{\ast} \mid w'\models \varphi, w'\in \{0,1\}^{n},\varphi\in \exists P_{n}\}
\]
Then it follows from the definition that $S=\bigcup_{n\in \mathbb{N}} S_{n}$ is in ${\bf NP}$. From ${\bf NP}\subseteq {\bf P}_{{\bf /poly}}$ we obtain a polynomial size sequence of propositional formulas $(\chi_{k})_{k\in \mathbb{N}}$ with $\chi_{k}$ deciding $S \cap \{0,1\}^{k}$.
To obtain the propositional formula equivalent to $\varphi=\exists q_{1}\cdots\exists q_{m}.\psi\in \exists P_{n}$
simply let $k$ be the length of $w_{\varphi}$ and consider $\chi_{k+n}$. Let the atoms of $\chi_{k+n}$ be $r_{1},\ldots,r_{k},p_{1},\ldots,p_{n}$ with $r_{i}$ used for the $i$-th component of $w_{\varphi}$. Then the formula obtained from $\chi_{k+n}$ by replacing $r_{i}$ by $\top$ if the $i$-th component of $w_{\varphi}$ is $1$ and by $\bot$ otherwise is as required.
This finishes the proof of Theorem~\ref{thm:uniformequiv}.

We next consider the size of Craig interpolants. Note that since uniform interpolants are also Craig interpolants, any lower bound on the size of Craig interpolants implies the same lower bound on the size of uniform interpolants. The following result is due to Mundici~\cite{DBLP:journals/apal/Mundici84}.
\begin{theorem}\label{thm:mundici}
    If propositional logic has polysize Craig interpolants, then ${\bf NP}_{{\bf /poly}}\cap {\bf coNP}_{{\bf /poly}}\subseteq {\bf P}_{{\bf /poly}}$,
    and so also ${\bf NP}\cap {\bf coNP} \subseteq {\bf P}_{{\bf /poly}}$.
\end{theorem}
\begin{proof}
Theorem~\ref{thm:inc} implies the ``also''-part. The main part is shown by using the definitions in a straightforward way. Assume propositional logic has polysize Craig interpolants. Also assume that $S\subseteq \{0,1\}^{\ast}$
is in ${\bf NP}_{{\bf /poly}}\cap {\bf coNP}_{{\bf /poly}}$. Then, since $S$ is in ${\bf NP}_{{\bf /poly}}\cap {\bf coNP}_{{\bf /poly}}$, we find families $(\exists {\bf q}_{n}.\varphi_{n})_{n\in \mathbb{N}}$ and $(\exists {\bf q}_{n}'.\varphi_{n}')_{n\in \mathbb{N}}$ of polysize deciding $S$ and $\{0,1\}^{\ast}\setminus S$, respectively.
Then, by definition,
$\exists {\bf q}_{n}.\varphi_{n} \models \neg \exists {\bf q}_{n}'.\varphi_{n}'$.
Hence $\varphi_{n} \models \neg \varphi_{n}'$
and so we find polysize Craig interpolants $\chi_{n}$ for $\varphi_{n},\neg\varphi_{n}'$. But then
$(\chi_{n})_{n\in \mathbb{N}}$ decides $S$ and so $S$ is in ${\bf P}_{{\bf /poly}}$.
\end{proof}
The question of whether the inclusion ${\bf NP}\cap {\bf coNP} \subseteq {\bf P}_{{\bf /poly}}$ holds is open, similar to the stronger inclusion ${\bf NP} \subseteq {\bf P}_{{\bf /poly}}$. Again, it is regarded very unlikely that the inclusion holds. However, if we could prove ${\bf NP}\cap {\bf coNP} \not\subseteq {\bf P}_{{\bf /poly}}$, then again ${\bf NP} \not= {\bf P}$ would follow. 

Further complexity-theoretic consequences of the assumption that propositional logic has polysize Craig interpolants are shown in~\cite{DBLP:conf/dagstuhl/SchoningT06}.
For instance, it would follow that ${\bf UP}\subseteq {\bf P}_{{\bf /poly}}$, where
${\bf UP}$ is the class of problems in ${\bf NP}$ accepted by non-deterministic Turing machines with at most one computation path on every input.

Interestingly, much more is known about the size of interpolants in the monotone fragment of propositional logic where only the connectives $\wedge$, $\vee$, $\top$, and $\bot$ are used. This fragment of propositional logic also has the Craig interpolation property, but now monotone formulas $\varphi,\psi$ are known for which Craig interpolants in the monotone fragment are of exponential size. We refer the reader to~\refchapter{chapter:proofcomplexity} for a discussion of the monotone fragment and also of \emph{feasible interpolation} where one does not measure the size of interpolants in the size of the input formulas but in the length of the proof of the entailment between them.

\section*{Acknowledgments}
\addcontentsline{toc}{section}{Acknowledgments}

The authors thank Jean Christoph Jung, Amirhossein Akbar Tabatabai as well as
Sam van Gool and his seminar students for their valuable comments and
suggestions. Funded by the Deutsche Forschungsgemeinschaft (DFG, German
Research Foundation) -- Project-ID~457292495.

\bibliography{bibexport_edited,taci}
\end{document}